\documentclass[a4paper,11pt]{article}
\pdfoutput=1 %
\usepackage{jheppub} 
\usepackage{subfigure}
\usepackage{subcaption}
\usepackage{graphicx}
\usepackage{amsmath}
\usepackage{physics}
\usepackage{tikz}
\usepackage{circuitikz}
\usepackage{slashed}
\usepackage{doi}
\usepackage{hyperref}

\newcommand{\ee}{{\mathrm e}}

\newcommand{\beq}{\begin{equation}}
\newcommand{\eeq}{\end{equation}}
\newcommand{\bea}{\begin{eqnarray}}
\newcommand{\eea}{\end{eqnarray}}

\newcommand{\Fig}[1]{Fig.~\ref{#1}}
\newcommand{\Eq}[1]{Eq.~(\ref{#1})}
\newcommand{\Sec}[1]{Sec.~\ref{#1}}
\newcommand{\App}[1]{Appendix~\ref{#1}}

\title{Two-point correlators in de Sitter-prepared states with bra-ket wormholes}

\date{\today}

\author[a,b]{Sunghoon Jung,}
\author[c,a]{Minju Kum,}
\author[a]{Junghwan Lee}
\affiliation[a]{Center for Theoretical Physics, Department of Physics and Astronomy, \\Seoul National University, Seoul 08826, Korea}
\affiliation[b]{Astronomy Research Center, Seoul National University, Seoul 08826, Korea}
\affiliation[c]{Department of Physics, University of Texas at Austin, Austin, TX 78712, USA}

\emailAdd{sunghoonj@snu.ac.kr}
\emailAdd{minjukum@utexas.edu}
\emailAdd{ghe1266@snu.ac.kr}

\abstract{
Motivated by the finiteness of de Sitter (dS) horizon entropy, we study how ``bra-ket wormholes'' modify correlation functions in gravitationally prepared states. Euclidean wormhole saddles in gravitational path integrals can generate non-factorizing contributions to correlation functions, as in replica-wormhole explanation of the Page curve and bra-ket-wormhole restoration of strong subadditivity. By defining `time' variables and computing observables in a flat region attached to the dS boundary, we evaluate bra–ket wormhole contributions to scalar two-point functions and find late-time transitions in the dominant saddle, accompanied by the ramp-and-plateau behavior of correlations and the characteristic timescale comparable to the fast scrambling. Each observable is consistent with `complementarity', in the sense that wormhole effects are distinguishable only at late respective times. Consistencies are based upon the interplay of (i) inflationary horizon exit and re-entry, (ii) enhancement of correlations at small comoving momentum by wormhole contributions, (iii) a competition between mode counting and topological suppression that drives a transition to wormhole dominance, which naturally yields the fast scrambling timescale, and (iv) irreducible errors by cosmic variance in early CMB-like observations. To clearly interpret in terms of entropy and chaotic nature of dS, one needs a more complete mechanism of wormhole stabilization.
}

\begin{document}
\maketitle
\setlength{\parskip}{0.5em}

\section{Introduction}

De Sitter(dS) spacetime possesses a cosmological horizon with Gibbons--Hawking thermodynamics \cite{Gibbons:1977mu}, associated with a finite entropy given by the Bekenstein-Hawking area law~\cite{Bekenstein:1973ur,Hawking:1975vcx}
\begin{equation}
S_{\mathrm{dS}} \,\sim\, \frac{A_{\mathrm{h}}}{4G\hbar},
\end{equation}
and---at least for suitably defined observables---a finite effective Hilbert space dimension,
\begin{equation}
\dim \mathcal{H} \,\sim\, e^{S_{\mathrm{dS}}}.
\end{equation}
If this interpretation is correct, then dS quantum gravity should exhibit the characteristic long-time constraints of finite-entropy systems \cite{Witten:2001kn,Goheer:2002vf,Parikh:2004wh}: information cannot be stored in an arbitrarily large number of independent degrees of freedom, and correlation functions cannot decay smoothly to zero forever~\cite{Maldacena:2001kr}.

This expectation is in direct tension with inflationary engineering. In slow-roll model building one can, within semiclassical effective field theory, tune the inflaton potential to realize an arbitrarily long quasi-dS stage with an arbitrarily large number of e-folds $N$. Each e-fold generates additional superhorizon modes that freeze out during inflation, implying an unbounded growth of cosmological information. This tension of dS is analogous to the black hole information problem.

In a post-reheating cosmology, this tension can be framed in terms of observables: a so-called CMB observer at post-reheating time $t$ can access only the subset of modes that have re-entered the horizon by that time, while the formal long-time limit $t\to\infty$ probes an ever larger set of inflationary modes. In parallel, one may regard $N$ not as dynamical time inside the dS path integral but as a \emph{state-preparation parameter} labeling a family of reheating-surface states $|\Psi(N)\rangle$. Varying $N$ provides a theoretical diagnostic of how observables depend on the inclusion of increasingly many horizon-exited modes---precisely where the clash with finite entropy is expected to sharpen. In this work, instead of $t$ and $N$, we will use $k_{\rm min}$ and $k_{\rm max}$ in the path-integral preparation of the state.

The gravitational path integral prepares a quantum state on the boundary of quasi-dS---reheating surface---by integrating over all possible past Euclidean geometries and field configurations given boundary conditions~\cite{Hartle:1983ai}. Although past time slices in the path integral cannot be interpreted as real dynamics of inflation and observables inside the dS are ambiguous~\cite{Witten:2001kn,Castro:2012gc}, the state can be unambiguously measured in a post-reheating flat region, providing a diagnosis of the dS quantum gravity. The two observables based on each notion of time allow complementary scrutinizations of the dS-prepared state.

In a finite-entropy system, late-time correlators are not expected to decay indefinitely; rather, they should exhibit residual correlations~\cite{Maldacena:2001kr} and, after appropriate averaging, ramp-and-plateau behavior~\cite{Saad:2018bqo,Saad:2019lba,Saad:2019pqd}. A natural scale for the onset of such deviations is a dS analogue of the Page time: irreducible errors due to the finite number of states $e^{-S_{\mathrm{dS}}}$ ($\propto e^{-1/G}$, hence non-perturbative) can become order one when enhanced by an effective multiplicity of available states of order $e^{S_{\mathrm{dS}}}$~\cite{Page:1993wv,Arkani-Hamed:2007ryv}. However, a horizon is also thought to be a chaotic fast-scrambling system, out of which information can leak far earlier than the Page time~\cite{Hayden:2007cs,Susskind:2011ap,Maldacena:2015waa}. Unlike the entanglement entropy, correlation functions can indeed be sensitive to wormhole effects at the scrambling time, e.g., via the scrambling shock-wave protocol~\cite{Shenker:2013pqa,Gao:2016bin,Maldacena:2017axo,Aalsma:2021kle}, although its generalization is unclear. Determining the mechanism that supplies these non-perturbative corrections and how it manifests in correlators of cosmological interest are the central motivations of this work.

A concrete precedent comes from the black hole information problem~\cite{Hawking:1975vcx,Preskill:1992tc}, where replica and bra-ket wormhole saddles modify semiclassical calculations and yield late-time behaviors consistent with unitarity~\cite{Almheiri:2020cfm,Penington:2019kki}. Inspired by this mechanism, we study whether analogous \emph{bra-ket wormholes} \cite{Chen:2020tes,Fumagalli:2024msi} contribute to correlators  in dS-prepared states, in ways consistent with finite-entropy constraints.

The paper is organized as follow. In \Sec{sec:GPS}, we introduce our calculation setup and the gravitationally prepared state with various consistency relations. In \Sec{sec:WH}, we find bra-ket wormhole saddle at the next-to-leading order(NLO) in topological expansion. In \Sec{sec:Obs-BKWH}, we calculate two-point correlators and discuss results of the two observables. Then we conclude with limitations and implications in \Sec{sec:discussion}.

\section{Gravitationally prepared state at the end of inflation}
\label{sec:GPS}

\begin{figure}
    \centering
    \includegraphics[width=0.5\linewidth]{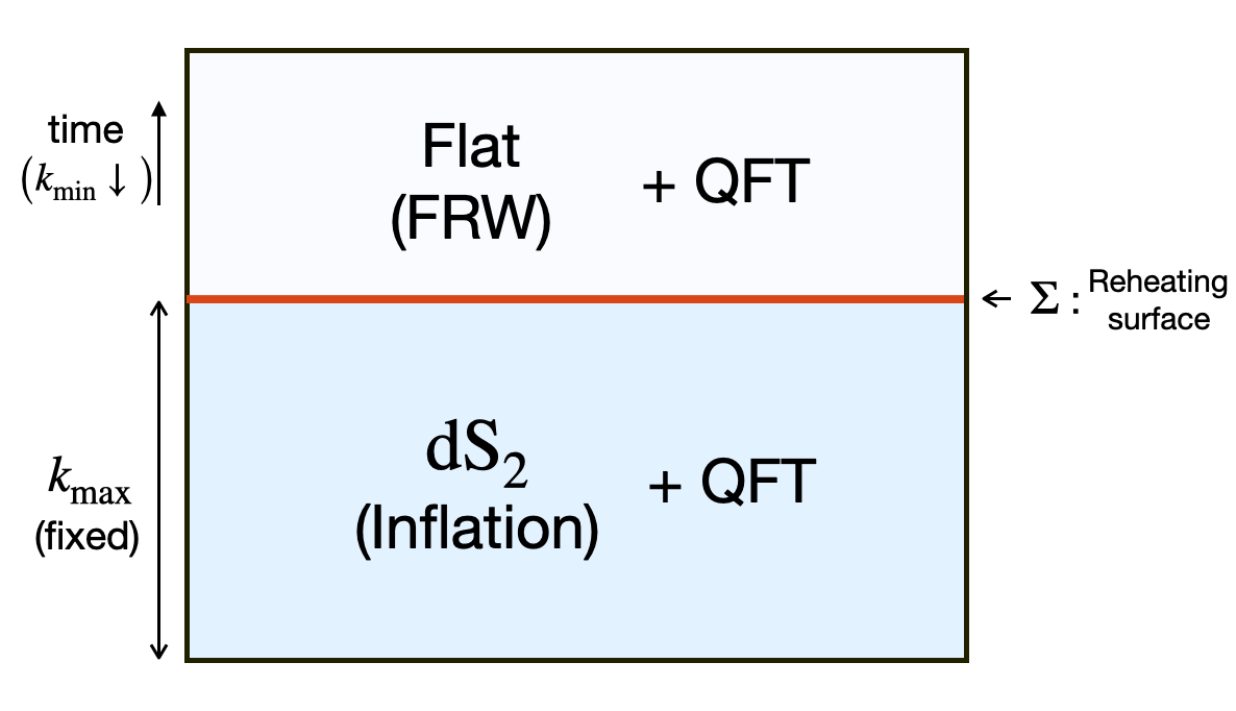}
    \caption{dS prepares a quantum state (of the scalar field) on its spatial boundary $\Sigma$, reheating surface, including effects from higher-topology wormholes. The state is then observable in the flat space afterwards without gravity. Two notions of time, $k_{\rm max}$ and $k_{\rm min}$, define two observables.}
    \label{fig:dSFRW}
\end{figure}

As depicted in \Fig{fig:dSFRW}, we gravitationally prepare a state on the future spatial boundary $\Sigma$ of dS space (\Sec{sec:stprep}), 
which is then observable in the flat space without gravity afterwards (\Sec{sec:stobs}). Thus, we assume finite dS, as a model of inflationary universe. On the contrary to black hole-AdS system, spacetimes with gravity turned on and off are timelike separated.

Gravitational path integral is said to gravitationally prepare a quantum state in the spatial boundary $\Sigma$
\beq
\int^\Sigma Dg D\Phi \, e^{iS},
\eeq
where geometry $g$ is fixed at $\Sigma$ but quantum fields $\Phi$ are left unspecified in general, so that the path integral is a functional of boundary field value $\phi = \Phi|_\Sigma$. The integral over $g$ can often be approximated by a discrete sum over saddle geometries $g_c$.

The action is given by the scalar quantum field theory on 2-dimensional JT \cite{Jackiw:1984je,Teitelboim:1983ux} dS space
\beq
    S \,=\, S_{\rm grav} \,+\, S_{QFT}[g,\Phi],
\label{eq:action0} \eeq
\beq
S_{\rm grav} \,=\, \frac{\varphi_0}{4G_2} \chi + \frac{1}{16 \pi G_2} \left( \int \sqrt{-g}\, \varphi (R-2) - 2 \int \sqrt{h}\, \varphi_b (K-1) \right).
\eeq
The dilaton $\varphi$ fixes $R=2$ by its equation of motion, its reference value is larger than the boundary value $\varphi_b \ll \varphi_0$, which determines the 2d dS entropy $S_{\rm dS} = 2\varphi_0/4G_2$. $\chi$ is the Euler characteristic (in the Lorentzian signature)
\beq
\chi \,=\, \frac{1}{4\pi} \left( \int \sqrt{-g}\, R - 2 \int \sqrt{h}\, K \right) \,=\, -i \chi_E, \qquad \chi_E \,=\, 2-2g-n,
\label{eq:chi}\eeq
providing topological expansion of geometries $e^{\frac{1}{2}S_{\rm dS} \chi_E}$.
The QFT action for a real scalar matter field is given by a minimal one 
\beq
S_{QFT}[g,\Phi]  \,=\, \frac{1}{2}\int \sqrt{-g}  \left(- g^{\mu\nu} \partial_\mu \Phi \partial_\nu \Phi - m^2 \Phi^2 \right).
\label{eq:actionm}\eeq

\subsection{Gravitational state preparation, and consistency} \label{sec:stprep}

To clarify notation and relative boundary conditions for bra and ket spaces, we rewrite gravitational state preparation while comparing with quantum mechanical bra-ket notation. Above all, we will consider only real fields on the boundary.

Define a gravitationally prepared ket state, in a given saddle geometry $g_c$,  (up to normalization)
\beq
U | 0 \rangle  \,=\,  \int^{}_{0} D\Phi \, e^{+i \int_0^t {\cal L}},
\label{eq:ketcontour} \eeq
where $0$ refers to a boundary condition defining a state (essentially selected by the saddle geometry $g_c$, in the Euclidean regime), and the boundary condition for the bra slice (where measurements are made) is open except that it is at time $t$. On the left-hand side, we use corresponding quantum mechanical bra-ket notation, where $U$ represents (both Euclidean and Lorentzian) path evolution of initial data $|0\rangle$ to the ket slice at $t$. When this ket state is projected onto the field basis $|\phi \rangle$ at $t$,
\beq
\langle \phi | U | 0 \rangle \,=\, \int^{\phi}_{0} D\Phi \, e^{+i \int_0^t {\cal L}}.
\eeq
This result might be interpreted as wavefunctional $\Psi[\phi]$.
The gravitationally prepared bra state is then related by complex conjugation
\beq
\langle 0 | U^\dagger \,=\, \int_{0*}^{} D\Phi e^{-i\int_{0*}^{t*} {\cal L}} \,=\, \int_{}^{0^*} D\Phi e^{+i \int_{t*}^{0*} {\cal L}},
\label{eq:bracontour} \eeq
using the elementary relation $(\int_a^b dx f(x))^* = F^*(b^*) - F^*(a^*) = \int_{a*}^{b*} dx f^*(x))$. The complex time variable is used to imply its analytic continuation. The latter form in \Eq{eq:bracontour} is useful when computing the elements of density matrix as one path integral
\beq
\rho \,\sim\, U|0\rangle \langle 0| U^\dagger \,\sim\, \int D\Phi e^{+i\int_{t*}^t {\cal L}}, \qquad
\rho_{\phi \phi'} \,\sim\, \int_{\phi^{'*}}^{\phi} D\Phi e^{+i\int_{t*}^t {\cal L}}.
\label{eq:rho_contour} \eeq
One takeaway message is: \emph{when bra and ket appear together, their boundary conditions are conjugated.} This is equivalent to having real eigenvalues of $\rho$, as you can check $\rho_{\phi \phi'} = \rho^*_{\phi' \phi}$ using the above path-integral representation. Among multiple bras, conjugation is not needed. 

In general, several saddle $|0\rangle$ (or $g_c$) contributions are summed in $\rho$. This is implicit in the notation of $|0\rangle$, which is actually a weighted superposition of various saddle contributions. In the path integral side, all possible past Euclidean saddles $g_c$ (with given bra/ket boundary conditions) are summed up according to path integral philosophy.

Here comes an important distinction of gravitational path integral from quantum mechanics. For saddle geometries that do not connect bra and ket boundaries, the contribution to $\rho$ is a product of bra and ket wavefunctions; this is same and clear in the quantum mechanical bra-ket notation. But path integral also allows wormhole geometries that connect bra and ket boundaries, which then cannot be factorized. This contribution may be subtle in terms of bra-ket notation of density matrix. And this wormhole contribution is what makes important deviation from mere quantum mechanical expectation based on unitarity and conservation of entropy, etc.

Expectation values of any observables ${\cal O}$ are calculable using gravitational path integral
\beq
\Tr [ {\cal O} \rho ] \,=\, \langle 0 | U^\dagger {\cal O} U | 0 \rangle \,\sim\, \int D\phi \int^\phi_{\phi^*} D\Phi \, {\cal O} \, e^{+i\int_{t^*}^t {\cal L}},
\label{eq:observable0} \eeq
where $D\phi = \Pi_k d \phi_k$ on the boundary fields.
Various wormhole contributions are implicit as discussed.
The two-point correlation of scalar matter field, which is our main interest, can be calculated by ${\cal O}=\phi(x)\phi(y)$ at the 1D spatial boundary at $t$.

\paragraph{Euclidean regime.}
The time contour can be analytically continued to the complex plane, given the fixed boundary real time.
Mathematically, it is a trick to do integral by finding out (complex) saddle geometries.

Physically, Euclidean regime is necessary to prepare states on the boundary. 
Without any imaginary parts in the time contour, the path integral is mere unitary evolution. Technically, the real part of action $S$ in $e^{iS}$ on the bra boundary is cancelled by that on the ket boundary; unitary evolution. Only the imaginary part of $S$ can lead to non-trivial states.

Not only time, but fields can also be complexified, just as another trick to do integral and solve EoMs. Complex fields can also provide extra contributions to states. But time's analytic continuation and fields' complexifications are not arbitrary, subject to several consistency conditions.

\paragraph{Probability interpretation.}
To interpret path integral as a partition function, one shall have a conserved probability. The gravitationally prepared state, being a solution of Wheeler-DeWitt (WDW) equation, has a conserved norm if the classicality condition is satisfied~\cite{Halliwell:1989ContNBWF,Hartle:2008ClassU} 
\beq
| \nabla_A {\rm Im}(S_c) | \, \ll \, |\nabla_A {\rm Re}(S_c) |
\label{eq:classicality}\eeq
for the action $S_c = S[g_c,\Phi_c]$ evaluated on a saddle \{$g_c, \Phi_c, \cdots$\}. $A$ represents the minisuperspace \{ $g, \Phi, \cdots$ \}. Then, the probability $P \approx |\Psi|^2 = e^{-2 {\rm Im}(S_c)}$ is conserved (and positive definite) along the trajectory $\nabla_A {\rm Re}(S_c)$. 
The classicality condition \Eq{eq:classicality} is reportedly equivalent to the reality condition on the boundary field values~\cite{Hartle:2008ClassU,Yeom:2014vba},
 \begin{equation}
     {\rm Im} (g_c) \ll {\rm Re} (g_c),\quad {\rm Im (\Phi_c)} \ll {\rm Re} (\Phi_c), \quad \cdots
 \label{eq:classicality2}\end{equation}
 
In all these, actions and fields are boundary quantities, expressed by boundary values in path integral. This is so because it is only the boundary state (the result of path integral) which becomes a member of WDW solutions, living on the minisuperspace.

Note that this approach of defining a state does not generally have a global consensus of time. This becomes particularly subtle in dS, as it does not have asymptotic flat regions.

\paragraph{Relaxed Kontsevich-Segal-Witten condition.}
For path integral to be convergent, the KSW condition requires the coefficient of field-quadratic term to be negative~\cite{Kontsevich:2021dmb,Witten:2021nzp}. Originally, this condition was required to be true on each point on the time contour. This is expressed as a condition on the metric components: $\sum_i |\arg(g_{ii})| < \pi$. For example, Lorentzian Minkowski space saturates this limit as $g_{tt} = -1 + \epsilon$ with the Feynman prescription $\epsilon >0$~\cite{Witten:2021nzp}. As we will see, our wormhole solution does not obey this strict KSW condition.

However, we suggest a relaxed condition, which requires only the result of path integral 
\begin{equation}
    \int^\phi D\Phi\,e^{iS[g_c,\Phi]} \,\propto\, e^{-I \phi^2}, \qquad {\rm Re}(I) \,>\, 0
\label{eq:relaxedKSW} \end{equation}
to be convergent, instead of requiring a minus sign point-wisely. Thus, Re($I$), obtained from the integral of the on-shell Lagrangian along the time contour, should be positive. We confirm this numerically for our wormhole solutions.

\paragraph{Hartle-Hawking geometry prepares the Bunch-Davies vacuum state.} 
The idea that a Euclidean geometry with right boundary condition determines the quantum state on the boundary (let alone the possibility of wormholes) was first and wonderfully realized by the no-boundary proposal by Hartle and Hawking~\cite{Hartle:1983ai}. This suggests that the leading contribution would be from a Euclidean hemisphere smoothly connected to the upper half of (the global coordinate of) the Lorentzian dS. This geometry is topologically leading order and is the most natural analytic continuation of the global coordinate. Most remarkably, the regularity of fields on the South pole successfully reproduces the unique Bunch-Davies vacuum on the Lorentzian part. See \Sec{sec:global} for more detailed introduction.

\subsection{Observation of dS prepared state, notion of time}  \label{sec:stobs}

{\bf Post-inflationary observation.} Our situation in \Fig{fig:dSFRW} is interpreted as inflation prepares a quantum state of the universe and the scalar field (representing the inflaton or the energy density fluctuation) on the reheating surface $\Sigma$, which is subsequently evolved into the post-inflationary FRW space. This state is observable by post-inflationary observers, e.g. by CMB and large scale structures.

As emphasized in \cite{Bousso:2000nf,Arkani-Hamed:2007ryv,Maldacena:2012xp,Teresi:2021qff}, this setup provides an operational or observational meaning to entropy, correlations, and other quantum properties of dS prepared state.

From the so called CMB observer's point of view, observed properties of dS state change with time, as successively smaller Fourier modes of quantum fields (larger wavelength) re-enter the horizon. The measured entropy also grows, and eventually becomes conflict with the finite dS entropy. Presumably, wormhole effects would cure this conflict, just like in the black hole case. In \Sec{sec:obs1}, we will show that wormhole effects encoded in various Fourier modes of dS state indeed make CMB observation at least qualitatively consistent with unitarity and complementarity.

{\bf Another notion of ``time''.} 
We are dealing with approximate dS, in which full isometries are slightly broken due to the end of inflation. This breaking naturally brings in the notion of ``time'' (during inflation), both as a cutoff theoretically and as duration phenomenologically. We emphasize that we do not study the dynamics of quantum states inside dS (during inflation). But the quantum state prepared at the end of inflation does depend on the cutoff or the total amount of inflation. By comparing quantum states of different universes with different duration, we might also be able to discuss some kind of time dependence of gravitationally prepared states. This also allows to discuss phase transition into wormhole dominance phase. As will be discussed in \Sec{sec:obs2}, this way of `observing' dS state is also consistent with resolving dS information problem.

\section{Wormholes for two-point correlations of scalar QFT} \label{sec:WH}

\subsection{Equations of motion and boundary conditions in the global coordinate} \label{sec:global}

We use the global coordinate in this work ($N=1$ minisuperspace with constant-time horizontal surfaces as foliation)
\beq
    ds^2 = -dt^2 + a(t)^2\, d\theta^2, \qquad \theta \in [0, 2\pi).
\label{eq:globalcoord} \eeq
To calculate QFT states on the future surface under dynamic gravity, we use gravitational path integral to evolve initial data (essentially selected by saddle geometry) on a Cauchy surface in the past Euclidean regime into future data on the physical boundary. To this end, the global coordinate is most flexible in analytically continuing the geometry from Lorentzian to Euclidean region. Both matter fields and the metric remain regular in this coordinate. The initial surface of Lorentzian dS is smoothly and naturally glued to the Euclidean region, just like the leading-order Hartle-Hawking(HH) no-boundary geometry.

In comparison, for CFT states (as often studied in literature), geometry plus CFT alone can determine the quantum state, while the path integral provides topological expansion of saddle geometries. Earlier works with CFT chose various coordinates (e.g. global in \cite{Teresi:2021qff} and Milne in \cite{Chen:2020tes,Fumagalli:2024msi}), as CFTs do not need to be solved analytically with full initial conditions. However, our goal is to obtain correlation functions of QFT matter fields, and for this we must solve EoMs in a foliation that covers both the entire initial surface and the late-time surface, at ease.

\paragraph{EoMs.} 
The variations of the gravity part of action \Eq{eq:action0} with respect to the dilaton $\varphi$ and metric field $g_{\mu\nu}$ yield the following EoMs,
\begin{equation}\label{eq:EoMformetric}
    R = 2, \quad (\nabla_\mu \nabla_\nu + g_{\mu\nu})\varphi \,=\, -8\pi G_2\,T_{\mu\nu}^M,
\end{equation}
with boundary conditions at the time \(t_\epsilon = \log (2/\epsilon)\) corresponding to the bra or ket space,
\begin{equation}\label{eq:bdycond_metric}
    ds_b^2 = \frac{d\theta^2}{\epsilon^2}, \quad \varphi_b = \frac{\varphi_r}{\epsilon}.
\end{equation}
The size of boundary is $\ell = 2\pi a(t_\epsilon) = 2\pi/\epsilon$.
The metric EoM reduces to
\begin{equation}
\ddot{a}(t) = \,a(t),
\end{equation}
which admits solutions expressed in terms of hyperbolic functions. 

In the given geometry $a(t)$, the bulk scalar field, $\Phi(t,\theta) = \sum_k e^{ik\theta} \Phi_k(t)$, obeys the following EoM from the matter action \Eq{eq:actionm}
\begin{equation}\label{eq:EoMscalar}
    \ddot{\Phi}_k(t) + \,\frac{\dot{a}(t)}{a(t)}\,\dot{\Phi}_k(t) + \left(\frac{k^2}{a(t)^2} + m^2\right) \Phi_k(t) = 0.
\end{equation}
At late time near boundary, the asymptotic form of solutions for \(\Phi_k(t)\) can be generally expressed as
\begin{equation}
    \Phi_k(t) \,\simeq\, \frac{\phi(k)}{u(t_\epsilon)}\,\left(\phi(k) \,a(t)^{-\Delta_-} + \mathcal O(k)\,a(t)^{-\Delta_+}\right), \qquad \Delta_\pm \,\equiv\, \frac{1}{2} \pm \nu, \quad \nu \,\equiv\, \sqrt{\frac{1}{4} - m^2},
\label{eq:Phi_asymp}\end{equation}
where $u(t_\epsilon) = \phi(k)\,\epsilon^{\Delta_-}+ \mathcal O(k) \epsilon^{\Delta_+}$ from the boundary condition $\Phi_k(t_\epsilon) = \phi(k)$.  $\mathcal O(k)$ is the piece that depends on saddle geometry $g_c$ and boundary conditions; thus, when saddle specific results are necessary, we will denote it by ${\cal O}^c(k)$ with the superscript $c$. But it is always ${\cal O}(k) \propto \phi(k)$. The on-shell action becomes a boundary term
\begin{equation}
S_{\rm QFT} \,=\, \pi \sum_k a(t) \Phi_{-k}(t) \partial_t \Phi_k(t) \Big|_{t_\epsilon} \,=\, -2\nu\pi\, a(t_\epsilon)^{\Delta_-} \sum_k \phi(-k) {\cal O}(k). 
\label{eq:gen_action} \end{equation}
Since ${\cal O}(k) \propto \phi(k)$, the action is quadratic in fields $S_{\rm QFT} \sim \sum_k {\cal A}_k \phi(k)\phi(-k)$. When bra and ket appear together in density matrix or two-point functions, only the imaginary part ${\rm Im}({\cal A}_k)$ survives in $i(S- S^*)$ (by the relative conjugacy of bra and ket boundary conditions) and determines two-point correlations.
This structure of solution is common to all dS geometries. 

The general structure that ${\cal O}(k)$ carries all saddle-dependent information and correlation information is in parallel with the AdS/CFT or dS/CFT correspondence. There, ${\cal O}(k)$ becomes a boundary CFT operator.
By QFT calculation in a non-trivial wormhole geometry, we explicitly check that this structure is still valid.

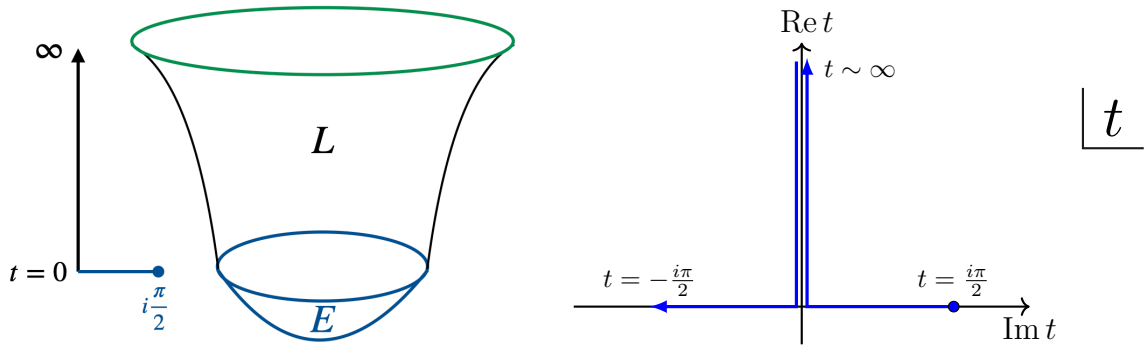
\begin{figure}
  \centering
  \subfigure{\includegraphics[width=0.45\linewidth]{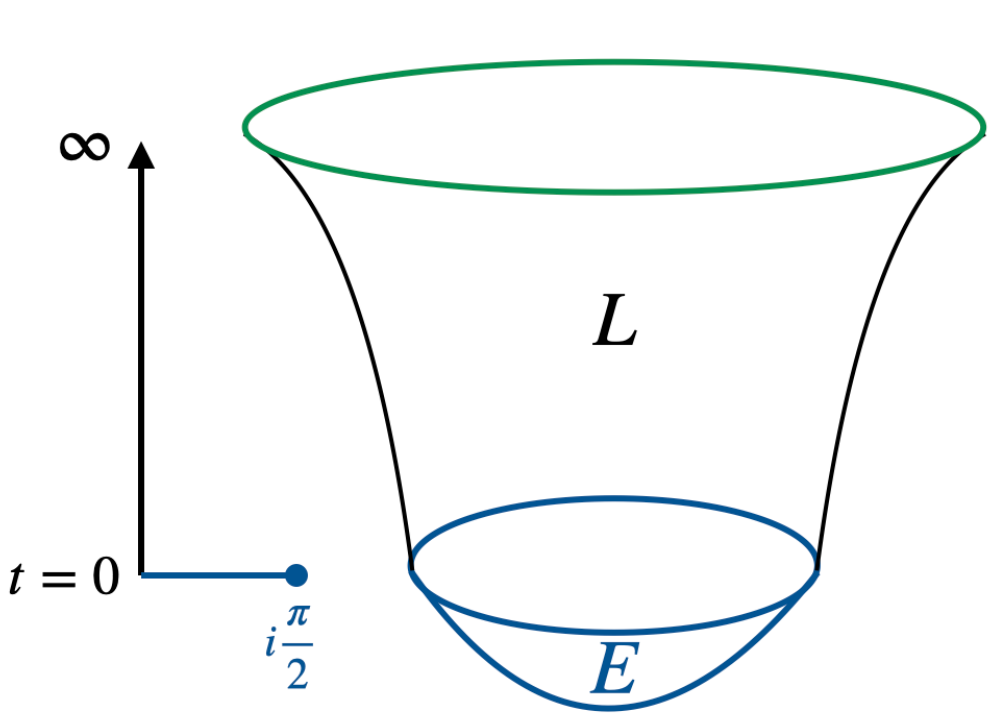}} 
  \hfill
  \subfigure{
    \begin{tikzpicture}
        \draw[->,thick] (-3,0) -- (3,0) node[below, font = \large] {${\rm Im\,}t$};
        \draw[->, thick] (0,-0.5) -- (0,3.5) node[above, font = \large, xshift=2] {${\rm Re\,}t$};
        \draw[->,>=latex, very thick, blue]  (2,0) -- (0.07,0) -- (0.07,3.3);
        \node[right,font=\small,black] at (0.16,3.15) {$t \sim \infty$} ;
        \draw[fill=blue,] (2,0) circle (2pt) node[align=center, above,black,font = \small] {$t = \frac{i\pi}{2}$};
        \draw[fill=blue,] (-0.07,3.75) circle;
        \draw[->,>=latex,very thick,blue] (-0.07,3.25) -- (-0.07,0) -- (-2,0);
        \node[align=center, above,black,font = \small] at (-2,0) {$t = -\frac{i\pi}{2}$};    
    
        \draw[black!90,thick] (3.7,2.9) |- (4.5,2.1) ;
        \node[black,scale=2] at (4.1,2.5) {$t$};
    \end{tikzpicture}
  }
  
  \caption{(Left:) HH no-boundary saddle in \Eq{eq:HHsaddle}. (Right:) Its contribution to the density matrix in \Fig{fig:SchematicDM} by a single path integral.}
  \label{fig:HH}
\end{figure}

\paragraph{Leading-order no-boundary saddle.}
As alluded, HH no-boundary geometry is a unique and natural saddle at LO. 
The geometry is given by 
\begin{align}\label{eq:HHsaddle}
    ds_{\text{Eucl.}}^2 &= d\tau^2 + \alpha^2 \cos^2(\tau)\,d\theta^2, \quad \tau \in [0,\pi/2], \\
    ds_{\text{Loren.}}^2 &= -dt^2 + \alpha^2 \cosh^2(t)\,d\theta^2, \quad t\in [0,\infty), \nonumber
\end{align}
as depicted in \Fig{fig:HH} left panel. The Euclidean hemisphere ($\chi_E=1$) is smoothly transitioned to the global coordinate at $t=0$. The South pole is located at $t=i\tau = i\pi/2$, at which fields are required to be regular. This natural boundary condition reproduces the Bunch-Davies vacuum on the future Lorentzian boundary. Exact solutions and on-shell actions are collected in \App{app:HH}. Its contribution to the density matrix can be calculated by a single path integral along the time contour shown in \Fig{fig:HH} right panel.

%

\subsection{Naive bra-ket wormholes do not exist at NLO} \label{sec:naiveWH}

\paragraph{Two-boundary connected geometry.} 
To find a bra-ket wormhole (in \Sec{sec:trace_univ}), we start with a two-boundary connected geometry. Depending on the nature of the two boundaries, this geometry can be called a bra-ket, bra-bra, or ket-ket wormhole. The topology at NLO is cylinder ($\chi_E=0$). 
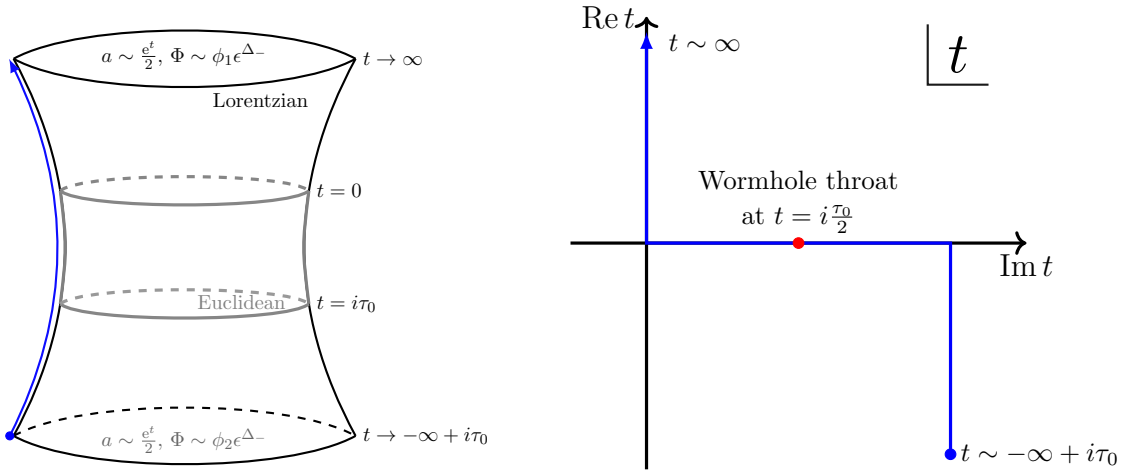
\begin{figure}[t]
  \centering
  \subfigure{
    \begin{tikzpicture}
    \draw[black, thick] (0,0) node[black, scale=0.7, left]{} .. controls (0.9,5/3) and (0.9,10/3) .. (0,5) node[black, scale=0.7, left]{};
    \draw[black, thick] (4.5,0) .. controls (3.6,5/3) and (3.6,10/3) .. (4.5,5);
    \draw[black,thick] (0,0) .. controls (0.9,-0.5) and (3.6,-0.5) .. (4.5,0) node[black,scale=0.7,right] {$t \to -\infty+i\tau_0$};
    \draw[black,thick,dashed] (0,0) .. controls (0.9,0.5) and (3.6,0.5) .. (4.5,0);
    \draw[black,thick] (0,5) .. controls (0.9,4.5) and (3.6,4.5) .. (4.5,5) node[black,scale=0.7,right] {$t \to \infty$};
    \draw[black,thick] (0,5) .. controls (0.9,5.5) and (3.6,5.5) .. (4.5,5);
    \draw[gray,very thick] (0.615, 1.75) .. controls (0.705,2.5) .. (0.615,3.25);
    \draw[gray,very thick] (3.885, 1.75) .. controls (3.795,2.5) .. (3.885,3.25);
    \draw[gray!95, very thick] (0.615, 1.75) .. controls (0.9,1.5) and (3.6,1.5) .. (3.885, 1.75) node[black,scale=0.7,right] {$t = i\tau_0$};
    \draw[gray!80, very thick,dashed] (0.615, 1.75) .. controls (0.9,2) and (3.6,2) .. (3.885, 1.75);
    \draw[gray!95, very thick] (0.615,3.25) .. controls (0.9,3) and (3.6,3) .. (3.885,3.25) node[black,scale=0.7,right] {$t = 0$};
    \draw[gray!95, very thick,dashed] (0.615,3.25) .. controls (0.9,3.5) and (3.6,3.5) .. (3.885,3.25);
    \draw[->,>=latex,blue!95, thick] (-0.05,0) .. controls (0.7875,5/3) and (0.7875,10/3) ..  (-0.06,5);
    \draw[fill=blue,color = blue] (-0.05,0) circle (1.5pt);
    \node[gray, scale = 0.7] at (3,1.77) {Euclidean};
    \node[black, scale = 0.7] at (3.25,4.45) {Lorentzian};
    \node[black!95, scale=0.7] at (2.25,5.05) {$a \sim \frac{\ee^t}{2},\,\Phi \sim \phi_1 \epsilon^{\Delta_-}$};
    \node[black!60, scale=0.7] at (2.25,-0.05) {$a \sim \frac{\ee^t}{2},\,\Phi \sim \phi_2 \epsilon^{\Delta_-}$};
    \end{tikzpicture}
    }
  \hfill
  \subfigure{
    \begin{tikzpicture}
        \draw[->,very thick] (-1,0) -- (5,0) node[below, font = \large] {${\rm Im\,}t$};
        \draw[->, very thick] (0,-3) -- (0,3) node[left, font = \large] {${\rm Re\,}t$};
        \draw[->,>=latex, very thick, blue] (4,-2.8) -- (4,0) -- (0,0) -- (0,2.8);
        \node[right,font=\small,black] at (0.16,2.65) {$t \sim \infty$} ;
        \draw[fill=blue,color=blue] (4,-2.8) circle (2pt) node[right,font = \small,black] {$t \sim -\infty + i\tau_0$};
        \draw[fill=red, red] (2,0) circle (2pt) node[align=center, above,black,font = \small] {Wormhole throat\\ at $t = i\frac{\tau_0}{2}$};
        \draw[black!90,thick] (3.7,2.9) |- (4.5,2.1) ;
        \node[black,scale=2] at (4.1,2.5) {$t$};
    \end{tikzpicture}
  }
  \caption{2-boundary wormhole saddle at NLO in \Eq{eq:bkwh0}, which is a building block of the bra-ket wormhole in \Fig{fig:SchematicDM} and \Sec{sec:trace_univ}. This by itself can only be a ket-ket or bra-bra wormhole because continuity of fields in the throat disallows proper boundary conditions for bra-ket wormholes (\Sec{sec:naiveWH}).}
  \label{fig:bkwh0}
\end{figure}

The only viable two-boundary solution for the scale factor is a complexified one given by
\begin{equation} \label{eq:bkwh0}
    a(t) = \frac{1}{2}\left(e^{t}+e^{i\tau_0-t}\right).
\end{equation}
The complex time contour connecting the two boundaries is depicted in \Fig{fig:bkwh0}. The geometry asymptotes to the real global coordinate $a(t) \to \cosh t$ toward each boundary $t\to \infty$ and $t \to -\infty + i\tau_0$. At $t=0$, a Euclidean cylinder is glued via analytic continuation $t= i\tau$ with $\tau \in [0,\tau_0]$, whose timelike length $\tau_0$ characterizes the Euclidean wormhole. 

Unlike usual analytic continuation of time variable alone, we complexify the functional form of $a(t)$ too (\Eq{eq:bkwh0}). The complexification is needed to have a solution. If two boundaries are simply connected through the periodic complex time contour $t= 0 \sim 2\pi i$, the geometry in the global coordinate becomes factorized into two HH hemispheres rather than a single connected bra-ket wormhole. This is different from the Milne coordinate and AdS, in which a doughnut-shaped wormhole along the periodic direction is obtained.

The complexified $a(t)$ may raise concerns. It is still imaginary along the real time line, and is not exactly real on the future cutoff surface at $t_\epsilon$. However, it still satisfies the classicality condition in \Eq{eq:classicality2} 
\begin{equation}
    a(t) \,\sim\, \dot a(t) \,\sim\, \tfrac12 e^t
\end{equation}
at late times $t \gg 1$, in that $a(t)$ and its conjugate momentum are almost real on the boundary. This was enough for probability interpretation.

The complex-valued scale factor induces additional complex phase, which only worsens the KSW condition and violates the condition at every point along the time contour. Nevertheless, the kernel resulting from the path integral remains positive so that the relaxed KSW is satisfied. We have checked this numerically.

The vacuum dilaton solution on this saddle geometry is
\begin{equation}
    \varphi(t) \,=\, \varphi_r\,e^{i\tau_0/2}\sinh\Bigl(t-\frac{i\tau_0}{2}\Bigr).
\end{equation}
Here, we neglect the contribution from the matter stress-energy tensor since it is exponentially suppressed at late times and subleading in \(G_2\).

The boundary extrinsic curvature on the ket side is given by
\begin{equation}
    K_{\rm ket} = 1 - \epsilon^2\left(\{f_{\rm ket}(\theta),\theta\} + e^{i\tau_0}\frac{f_{\rm ket}'(\theta)^2}{2}\right),
\end{equation}
with the bra side similarly by complex conjugation. The corresponding on-shell action is
\begin{align}\label{eq:Sgrav-WH}
   iS_{\rm grav}&= \frac{i\varphi_r}{8\pi G}\int_0^{2\pi}d\theta\,\Biggl[\Bigl(\{f_{\rm ket}(\theta),\theta\} + e^{i\tau_0}\frac{f_{\rm ket}'(\theta)^2}{2}\Bigr) - \Bigl(\{f_{\rm bra}(\theta),\theta\} + e^{-i\tau_0}\frac{f_{\rm bra}'(\theta)^2}{2}\Bigr)\Biggr] \nonumber\\[1mm]
    &= -\frac{\varphi_r}{4G}\sin\tau_0,
\end{align}
where in the final line \(f_{\rm ket}(\theta)=f_{\rm bra}(\theta)=\theta\) was used.\footnote{Quantum corrections from fluctuations of the Schwarzian mode \(f(\theta)=\theta_0+\delta\theta\) are known exactly \cite{Stanford:2017thb,Maldacena:2019cbz}, but are not included here.}

Note that the gravity part $S_{\rm grav}$ is not exactly real on the boundary so that the gravity contribution is not cancelled between bra and ket sides in \Eq{eq:Sgrav-WH}. This is due to the complexified $a(t)$. This is anyway consistent as discussed, and is also numerically small since $\varphi_r \ll \varphi_0$.

\paragraph{Matter solution.} 
For QFT scalar matter fields, we impose Dirichlet boundary conditions on both boundaries at $t = t_\epsilon,\, -t_\epsilon + i \tau_0$,
\begin{equation}\label{eq:Ewh BC1}
    \Phi(t_\epsilon,k) =\phi_1(k),~~ \Phi(-t_\epsilon+i\tau_0,k) = \phi_2(k). 
\end{equation}
A solution to the scalar field EoM can be expressed similarly to the HH solution,
\begin{equation}\label{eq:Ewh BC 2}
    \Phi(t,k) = \phi_1(k) \frac{\upsilon_k(t)}{\upsilon_k(t_\epsilon)} = \phi_2(k) \frac{\upsilon_k(t)}{\upsilon_k(-t_\epsilon + i\tau_0)}.
\end{equation}
$\upsilon_k(t)$ is given in \App{app:field_sol}. By employing the asymptotic form of solutions in \Eq{eq:Phi_asymp}
\begin{equation}
    \upsilon_k(t) \,\simeq\,\phi(k) \, a(t)^{-\Delta_-} + \mathcal O^{WH}(k) \,a(t)^{-\Delta_+}, 
\label{eq:asymp-WH}\end{equation}
the matter action is evaluated as (\Eq{eq:gen_action})
\bea
     i S_{\rm QFT} [g_c,\Phi_{c}] &\,=\,& i\pi \sum_k a_c(t) \partial_t\Phi_{c}(t,k) \Phi_{c}(t,-k) \Big|_{-t_\epsilon+i\tau_0}^{t_\epsilon}\\\nonumber
    &\,=\,& -2i\pi \nu \sum_k  \left(\frac{2\pi}{\ell}\right)^{-\Delta_-}\Big[\phi_1(-k) \mathcal{O}^{WH}_1(k)  + \phi_2(-k) \mathcal{O}^{WH}_2(k)\Big].
\label{eq:Sqft1} \eea
On each boundary, say 1, $\mathcal O_1^{WH} = \tilde a \phi_1 + \tilde b\phi_2$, where $\tilde b$ parametrizes the mixing between two boundary field values $\phi_1$ and $\phi_2$. Consequently, the resulting action is quadratic in the boundary values $\phi_{1,2}$. This mixing is critical in understanding various properties of wormholes, as will be discussed throughout.

Although complex in bulk, $\Phi$ also obeys the classicality condition. From the asymptotic form in \Eq{eq:asymp-WH}, the only piece that might contain sizable imaginary part on the boundary (hence, problematic) is ${\cal O}^{WH}(k)$. But this second term is exponentially smaller than the first term. Likewise, its conjugate momentum brings only $\Delta_{\pm}$ factors to the front, which is not of exponential in size. Thus, $\Phi$ and its conjugate momentum are almost real on the boundary.

\paragraph{This is not a bra-ket wormhole.} 

As discussed in \Sec{sec:stprep}, for the two boundaries to be identified as a bra and a ket (thus forming a bra-ket wormhole), bulk fields on either side of wormhole throat was required to be complex conjugate of each other. But our two-boundary wormhole above does not satisfy this boundary condition. This can be checked directly or as following.

The shift $t\to -t + i\tau_0$ maps the bulk on one side to the other side in the wormhole geometry. The explicit solution that we found is symmetric under this without complex conjugation
\beq
a(t) \,=\, a(-t + i\tau_0).
\eeq
The matter field saddle inherits the same symmetry, since its EoM is determined by this background. Therefore, this two-boundary geometry cannot be a bra-ket wormhole.
But this geometry can still be a bra-bra or a ket-ket wormhole. In the next subsection, we combine these bra-bra and ket-ket wormholes to construct a bra-ket wormhole.

\subsection{Bra-ket wormhole by tracing out unobservable universe}\label{sec:trace_univ}

\begin{figure}
    \centering
    \includegraphics[width=0.9\linewidth]{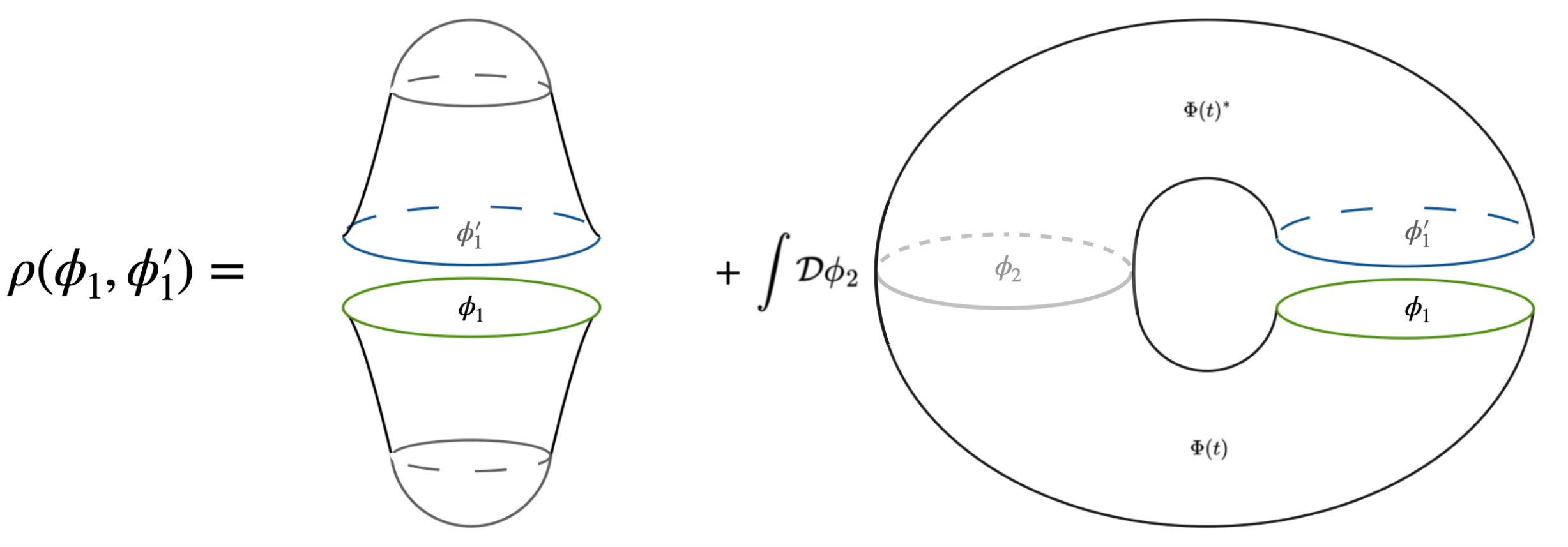}
    \caption{Gravitational path integral calculation of density matrix, in the field basis. The leading contribution is a product of HH states, while the NLO contribution is the effective bra-ket wormhole, which is obtained by joining two 2-boundary wormholes in \Fig{fig:bkwh0} and tracing out unobservable separate universe (denoted with $\phi_2$); see \Sec{sec:trace_univ}.
    }
    \label{fig:SchematicDM}
\end{figure}

We construct an effective bra-ket wormhole by combining a bra-bra and a ket-ket wormhole,  as depicted in \Fig{fig:SchematicDM}. The upper half, which is a ket-ket wormhole, is a complex conjugate of the lower half, which is a bra-bra wormhole, so that the bra-ket relative conjugacy condition is effectively satisfied. Each bra-bra or ket-ket wormhole is thought to produce two separate universes, with independent boundary conditions $\phi_1$ and $\phi_2$. Then the unobservable universes (say with $\phi_2$) from bra-bra and ket-ket wormholes are joined and traced over. This results in a proper bra-ket wormhole, that contributes to the desired density matrix 
\begin{equation}
    \rho[\phi_1,\phi'_1] \,=\, \int \mathcal D \phi_2 \,\rho[\{\phi_1,\phi_2\},\emptyset]\,\rho[\emptyset,\{\phi'_1,\phi_2\}].
\end{equation}
This idea was also suggested by \cite{Anous:2020dmu} in a situation where direct bra-ket wormholes did not exist.

Let us compute more concretely. The path integral over each bra-bra and ket-ket wormhole results in each density matrix
\begin{equation}
     \int_{\Phi_\partial =(\phi_1,\phi_2)} \mathcal D\Phi\mathcal Dg\,e^{iS[g,\Phi]} \simeq e^{iS_c[g_c,\Phi_c]},
\end{equation}
where the on-shell action evaluated on the saddle $(g_c,\Phi_c)$ is in the form of
\begin{align}
    i S_c[g_c,\Phi_c] &\,=\, i\pi \sum_k a_c(t) \partial_t\Phi_{c}(t,k) \Phi_{c}(t,-k) \Big|_{-t_c+i\tau_0}^{t_c}\\\nonumber
    &= -\sum_k \Bigl\{ a_k \phi_1(k) \phi_1(-k) - \bigl[b_k \phi_1(k)\phi_2(-k) + c.c. \bigr] +a_k \phi_2(k)\phi_2(-k) \Bigr\}.
\end{align}
Here, $a_k,b_k$ are the constants determined by bulk scalar solution $\Phi_c$; see \App{app:WH} for exact expressions. By combining bra-bra and ket-ket states, i.e. $iS_c[\Phi_c] -iS_c^*[\Phi_c^*]$, the total density matrix is in the form of (for each $k$-mode)
\begin{equation}
    \int \mathcal D\phi_2\,\exp\Bigl[ -\alpha_k |\phi_1|^2 + \beta_k(\phi_1^*\phi_2 + \phi_1 \phi_2^*) - \alpha_k |\phi_2|^2  \Bigr] \,\propto\, \exp\Bigl[-\left(\alpha_k-\tfrac{\beta_k^2}{\alpha_k}\Bigr) |\phi_1|^2\right],
\label{eq:braketS} \end{equation}
where $\alpha_k= 2{\rm Re}(a_k)$ and $\beta_k=2 {\rm Re}(b_k)$. This is for each $k$-mode. As discussed, the action is quadratic in field, and its coefficient leads to the two-point correlation at leading order.

\paragraph{Consistencies.}
This geometry obeys the classicality condition, simply because each bra-bra and ket-ket does so. 
The strict KSW is still not obeyed, but we checked numerically that the relaxed KSW is satisfied. This means that the coefficient of quadratic term for each $k$ is positive: $\alpha_k > \beta_k$ for all $k$, from \Eq{eq:braketS}.

Moreover, for large $k > k^*$, the wormhole action in \Eq{eq:braketS} agrees with that of HH. This must be so because modes that are shorter than the wormhole scale ($\tau_0$) or the curvature of space are insensitive to the global geometry. This means that $\beta_k \to 0$ for large $k>k^*$.

As a result of these consistencies, the effective coefficient of quadratic term is reduced for small $k$: $\alpha_k \to \alpha_k - \beta_k^2/\alpha_k$ from \Eq{eq:braketS}. This enhances low-$k$ (long-range) correlations. As will be discussed in \Sec{sec:mech}, this enhancement will be critical in phase transition to wormhole dominance.

\paragraph{Never ending (bra-bra)+(ket-ket) chains?}
One may wonder if our mechanism of forming an effective bra-ket wormhole by joining bra-bra and ket-ket wormholes can be extended by joining even more chains of bra-bra and ket-ket wormholes. They are indeed legible contributions. But, as detailed in \App{app:Magnet}, adding $n$ such chains increases the action in proportion to $\propto n$ so that they are exponentially suppressed. Creating longer and more complex wormholes essentially costs more resources. We ignore these contributions.

\subsection{Stabilization of wormholes} \label{sec:stabilization}

The partition function from each $k$ mode 
\begin{equation}
    \int D\phi_k e^{-a_k \phi_k^2} =  \sqrt{\frac{1}{a_k}} = e^{-\frac{1}{2}\log a_k},
\end{equation}
is diverging as $\tau_0 \to 0$ since $a_k \simeq c_k \tau_0$ for small $\tau_0$. 
In this $\tau_0 \to 0$ limit (equivalently, in the limit of infinite temperature), the partition function corresponds to merely counting total number of states, as Boltzmann suppression of high-energy states is absent. In the finite-entropy system, this must be finite and, moreover, bounded by about $e^{S_{dS}}$. So the divergence sounds like a pathology of semiclassical effective theory. Further, this problem is related to the stabilization of the wormhole; $\tau_0 =0$ means that the size of Euclidean geometry vanishes. If wormholes are to make sizable contributions, $\tau_0$ shall be stabilized to some non-zero finite value. The wormhole stabilization is a more general problem in dS, even with CFT matter in the bulk~\cite{Fumagalli:2024msi,Chen:2020tes,Milekhin:2022yzb}. 

In the random state model of quantum gravity~\cite{Jung:2025rip}, wormholes indeed turn out to be stabilized; without encountering any divergences, this model successfully reproduces the expected properties of correlation functions in the wormhole dominant phases. Also, in Ref.~\cite{Fumagalli:2024msi}, Wigner distribution of observables also provided some stabilizing force. Accepting these general expectations, we will simply assume a non-zero finite value of $\tau_0 = 0.01$ in our numerical calculations. But true saddle points shall be in relation to $S_{\rm dS}$ as will be discussed, which we leave for future works.

Naively, the divergence at $\tau_0 \to 0$ seems to be regularized by $\zeta$ function
\bea\label{eq:stab-scalar}
    \int D\phi\,e^{-\sum_k a_k \phi_k^2} &\,=\,& \prod_k \sqrt{\frac{\pi}{a_k}} \\
    &\,\to\,& \exp \left[ \tfrac12 \left(2 \sum_{k\in \mathbb N}1+1 \right) \log\tau_0 + \cdots \right] \,=\, \exp( \tfrac{1}{2} (2 \zeta(0) + 1) \log \tau_0 + \cdots) \nonumber
\eea
where the ellipsis indicates $\tau_0$–independent finite constants. This is valid for small $\tau_0$ (where $a_k \propto \tau_0$ linearly). One can regulate the sum over natural numbers using the $\zeta$ function with  $\zeta(0) = -\tfrac12$, and this exactly cancels the $\log \tau_0$ divergence. However, this infinite sum is not  consistent with the cutoff $k_{\rm max}$ of semiclassical effective theory, which will be introduced and utilized in our observables. Thus, we do not take this possibility seriously.

\paragraph{Fermions added.} We remark that both problems can be explicitly resolved by introducing a Majorana fermion $\psi$ (two real components) for each complex scalar field $\phi$. The Gaussian integral over anticommuting fields leads to the opposite $\tau_0$ dependence
\begin{equation}
\int D\psi_k D\bar{\psi}_k e^{-b_k \bar{\psi}_k \psi_k} = b_k = e^{+\log b_k},
\label{eq:stab-fermion}\end{equation}
where $b_k \sim \tau_0$ as $\tau_0 \to 0$. The detailed calculation is described in \App{app:Fermi-Stab}. 
Thus, for each mode $k$, the scalar ($\sim e^{-\tau_0}$) and fermion ($\sim e^{+\tau_0}$) contributions cancel the leading $\log \tau_0$ divergence. In a sense, fermions repel wormholes to shrink completely.
The residual dependence on $\tau_0$ also leads to non-zero finite saddle value of $\tau_0$, essentially due to the same opposite dependence. An example result in \Fig{Fig:ZratSF03} shows that the saddle is located at $\tau_0 \simeq 10^{-3}$. 

However, this mechanism of stabilization cannot be complete as it does not depend on the gravity action. Thus, we do not take this resolution with extra fermions seriously. Rather, in this study, we simply take the fixed value $\tau_0 = 0.01$.

\section{Observables of bra-ket wormholes}\label{sec:Obs-BKWH}

\subsection{Notions of time: $k_{\rm max}$ and $k_{\rm min}$}

In post-inflationary observation of dS prepared state, only those modes of scalar field that became superhorizon by the end of inflation will be classicalized and observable. We introduce upper limit on the observable Fourier modes of scalar field, $k_{\rm max}$, as the largest $k$-mode that existed the horizon ($1/H$) by the end of inflation
\beq
\frac{2\pi a(t_\epsilon)}{H k_{\rm max} } \,=\, \frac{1}{H} \quad \to \quad k_{\rm max} \,=\, 2\pi a(t_\epsilon).
\label{eq:kmaxdef} \eeq
$k_{\rm max}$ appears in path integral calculation of correlation functions in the following way 
\bea
\langle \phi_k \phi_{-k} \rangle &\,=\,& \frac{ \int^{k_{\rm max}} \mathcal{D}\phi \,\phi_k \phi_{-k}\,\left\{ \, e^{S_{\rm dS}} \exp\Bigl(-4\pi\sum_{q}^{k_{\rm max}}{\rm Im}\,\mathcal{A}_q^{\rm HH}\,\phi_q\phi_{-q}\Bigr)
    + \left. \exp\Bigl(-4\pi\sum_{q}^{k_{\rm max}}{\rm Im}\,\mathcal{A}_q^{\rm WH}\,\phi_q\phi_{-q}\Bigr)\right|_{\tau_0} \right\} }
    {\int^{k_{\rm max}} \mathcal{D}\phi\,\left\{ \,e^{S_{\rm dS}}\exp\Bigl(-4\pi\sum_{q}^{k_{\rm max}}{\rm Im}\,\mathcal{A}_q^{\rm HH}\,\phi_q\phi_{-q}\Bigr)
    + \left. \exp\Bigl(-4\pi\sum_{q}^{k_{\rm max}}{\rm Im}\,\mathcal{A}_q^{\rm WH}\,\phi_q\phi_{-q}\Bigr) \right|_{\tau_0} \right\} } \nonumber\\[1mm]
&\,=\,& \frac{\frac{1}{8\pi\,{\rm Im}\,\mathcal{A}_k^{\rm HH}}\prod_{q}^{k_{\rm max}}\sqrt{\frac{\pi}{4\pi\,{\rm Im}\,\mathcal{A}_q^{\rm HH}}}
    + e^{-S_{\rm dS}}\frac{1}{8\pi\,{\rm Im}\,\mathcal{A}_k^{\rm WH}(\tau_0)}\prod_{q}^{k_{\rm max}}\sqrt{\frac{\pi}{4\pi\,{\rm Im}\,\mathcal{A}_q^{\rm WH}(\tau_0)}}
    }{\prod_{q}^{k_{\rm max}}\sqrt{\frac{\pi}{4\pi\,{\rm Im}\,\mathcal{A}_q^{\rm HH}}}
    +  e^{-S_{\rm dS}} \prod_{q}^{k_{\rm max}}\sqrt{\frac{\pi}{4\pi\,{\rm Im}\,\mathcal{A}_q^{\rm WH}(\tau_0)}} }. 
\label{eq:obs1} \eea
Here, ${\cal D}\phi = \Pi_k d\phi_k $ on boundary modes.
We sum only observable superhorizon modes $k \leq k_{\rm max}$ both in the path-integral measure and in the action sum. 
Subhorizon modes are quantum corrections, affecting correlations of superhorizon modes only through loop effects. These are not expected to be significant after inflation ends. 

Based on \Eq{eq:obs1}, Observable 1 considers two-point correlations in the dS prepared state with fixed $k_{\rm max}$: $\langle \phi_k \phi_{-k} \rangle$ for $0\leq k \leq k_{\rm max}$ and its spatial correlation 
\beq
\langle \phi(\theta) \phi(0) \rangle \,=\, \sum_{k \in \mathbb Z}^{k_{\rm max}} e^{i k \theta} \langle \phi_k \phi_{-k} \rangle.
\label{eq:spcorr1} \eeq
By $k_{\rm max}$, we do not mean to study dynamics during inflation (inside dS); $k_{\rm max}$ is a fixed quantity for each dS universe. However, the quantum state at the end of inflation does depend on $k_{\rm max}$. Thus, varying $k_{\rm max}$ means comparing states from different universes with different duration of inflation. In this way, we can discuss some kind of time dependence, such as when the phase transition to wormhole dominance occurs.

$k_{\rm max}$ happens to be the same as the cutoff of dS space: $k_{\rm max} = 2\pi/\epsilon = 2\pi a(t_\epsilon)$ from \Eq{eq:bdycond_metric}. The introduction of $\epsilon$ did not force to introduce $k_{\rm max}$. It is inflationary physics that requires to introduce $k_{\rm max}$ and provides absolute physical meaning.

Observable 2, on the other hand, considers CMB-like two-point correlations, hence only among those modes that have successively re-entered the horizon by given time. This introduces another notion of time: $k_{\rm min}$ as the smallest Fourier modes of scalar fields that have re-entered by given time 
\beq
\langle \phi(\theta) \phi(0) \rangle \,=\, \sum_{k = k_{\rm min}}^{k_{\rm max}} e^{i k \theta} \langle \phi_k \phi_{-k} \rangle,
\label{eq:spcorr2} \eeq
with $\langle \phi_k \phi_{-k} \rangle$ given by \Eq{eq:obs1}.
The farther future from reheating, the smaller $k_{\rm min}$ and the more modes have re-entered the horizon. By varying $k_{\rm min}$, we really mean to study the time-dependence of CMB observation of the given prepared state by a single observer. 

$k_{\rm min}$ appears only in the Fourier sum in \Eq{eq:spcorr2}, but not in the path integral measure and the action sum in \Eq{eq:obs1}. The state---particular set of Fourier modes with  weights---is already fixed by $k_{\rm max}$-path integral. It is only a matter of which modes among them to combine for correlation observables.

In summary, the two notions of time provide complementary diagnoses of the dS prepared state; one as a proxy of the time-dependence of dS gravitational non-perturbative effects, and the other as a CMB observation clock. Each observable shall make sense by itself, as some kind of dS complementarity.

\subsection{Phase transition to wormhole dominance} \label{sec:mech}

If wormholes are to resolve information problem, they shall presumably begin to dominate at some late stage of dS. How does this phase transition between different dominant saddles occur in our solution? We discuss its basic mechanism, which will be realized slightly differently in our two observables.

See \Eq{eq:obs1} second line numerator. Which saddle dominates is essentially a question of which of the following term is larger
\beq
\prod_q^{k_{\rm max}} \sqrt{\frac{1}{{\rm Im} {\cal A}_q^{HH}} }\qquad {\rm vs.} \qquad e^{-S_{\rm dS}} \cdot \prod_q^{k_{\rm max}} \sqrt{ \frac{1}{{\rm Im}{\cal A}_q^{WH}} }.
\label{eq:phasecompet} \eeq
For the latter wormhole effect to dominate, it is critical to have $1/{\rm Im}{\cal A}_k^{HH} < 1/{\rm Im}{\cal A}_k^{WH}$. Then, by having many enough modes (with some large enough $k_{\rm max}$), the multiplication of $1/{\rm Im}{\cal A}_q^{WH}$ terms can overcome topological suppression $e^{-S_{\rm dS}}$. If true, in other words, only long enough inflation would ever experience the phase of wormhole dominance. This kind of competition between the number of modes versus topological suppression, at least qualitatively agrees with the general mechanism of resolving information problem via connected geometries~\cite{Penington:2019kki, Almheiri:2020cfm}. 

As a result, depending on the ratio $Z_{\rm rat} \equiv \prod_q^{k_{\rm max}} \sqrt{ \frac{{\rm Im} {\cal A}_q^{HH}}{{\rm Im} {\cal A}_q^{WH}} }$ vs. $e^{-S_{\rm dS}}$,
\bea
    \langle \phi_k\phi_{-k}\rangle &\,=\,& \frac{1}{1+e^{-S_{\rm dS}}Z_{\rm rat}}\frac{1}{8\pi\,{\rm Im}\,\mathcal{A}_k^{\rm HH}}+\frac{e^{-S_{\rm dS}}Z_{\rm rat}}{1+e^{-S_{\rm dS}}Z_{\rm rat}}\frac{1}{8\pi\,{\rm Im}\,\mathcal{A}_k^{\rm WH}} \nonumber\\[2mm]
    &\,\simeq \,&
    \begin{cases}
    \displaystyle \frac{1}{8\pi\,{\rm Im}\,\mathcal{A}_k^{\rm HH}}, & \quad Z_{\rm rat}\ll e^{S_{\rm dS}}, \\[2mm]
    \displaystyle \frac{1}{8\pi\,{\rm Im}\,\mathcal{A}_k^{\rm WH}}, & \quad Z_{\rm rat}\gg e^{S_{\rm dS}}.
    \end{cases}
\label{eq:obs1-2}\eea
We denote the critical point by $k_{\rm max}^c$, where $Z_{\rm rat} = e^{-S_{\rm dS}}$. This is where phase transition occurs in the full state, or in Observable 1. When we say phase transition, we usually refer to this one. Another phase transition in Observable 2, $k_{\rm min}^c$, will also be introduced later, based on the same mechanism while realized slightly differently.

In our bra-ket wormhole solution, the necessary condition ${\cal A}_k^{WH} < {\cal A}_k^{HH}$ is satisfied for small $k<k^*$ as discussed in \Sec{sec:trace_univ}; for large $k>k^*$, WH contributions agree with HH. Briefly recapping, bra-bra wormhole-induced mixing (or entanglement) between scalar fields in separate universes
\begin{equation}
-a_k \phi_1^2(k) + 2b_k \phi_1(k) \phi_2(k) -a_k\phi_2^2(k)  \quad \to\quad -\Bigl(a_k-\frac{b_k^2}{a_k}\Bigr) \phi_1^2(k)
\end{equation}
effectively enhances low-$k$ mode correlations $\sim 1/a_k^{\rm eff}$ of $\phi_1$, by effectively reducing $a_k \to a_k^{\rm eff} = (a_k^2 - b_k^2)/a_k$. For this to work, $b_k < a_k$, which is consistent with the relaxed KSW.

The resulting enhancement of low-$k$ correlations are numerically confirmed in later sections. This is necessary for the KSW as well as phase transitions into wormhole dominance. However, whether and how these are due to the structure of our bra-ket wormhole (tracing out the second universe) is not clear to us. At least, tracing out the half of the dS boundary space is known to induce such an enhancement of low-$k$ correlations in the remaining subsystem~\cite{Kanno:2014ifa}. Wormholes may induce entangled states, e.g. as in ER=EPR conjecture~\cite{Maldacena:2013xja,Susskind:2014yaa}, or in another way, interactions may induce wormholes~\cite{Gao:2016bin,Maldacena:2017axo,Maldacena:2018lmt} or entangled states~\cite{Cottrell:2018ash}. The scrambling physics that will be discussed in next subsections may also be relevant.

\subsection{Parameter space: Scaling and Scrambling} \label{sec:paramsp}

\begin{figure}[t]
    \centering
    \includegraphics[width=0.7\linewidth]{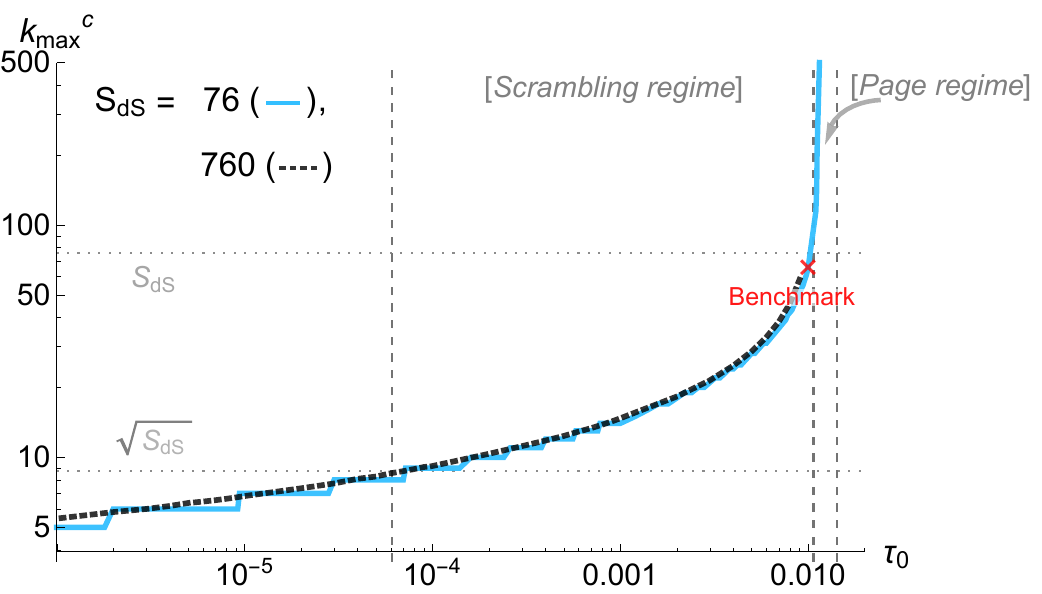}
    \caption{The scaling relation between free parameters of the model, $\{ S_{\rm dS}, \tau_0, k_{\rm max} \}$, using $k_{\rm max}^c$ as a function of $\tau_0$ for two choices of $S_{\rm dS}=76$(solid), 760(dashed). Shown numerical values are for $S_{\rm dS}=76$, while the result of $S_{\rm dS}=760$ is rescaled by \Eq{eq:scaling} with $c=1/10$, and they overlap very well. Furthermore, the scaling allows to categorize the whole parameter space into three regimes separated by vertical lines. The scrambling timescale turns out to be relevant to the bra-ket wormhole, but the exact location of a theory depends on a true stabilization mechanism. The benchmark parameter in the scrambling regime is marked as a cross.}
    \label{fig:scaling}
\end{figure}

Our model has four free parameters: $\{ S_{\rm dS}, \tau_0, k_{\rm max}, m \}$. For Observable 2, $k_{\rm min}$ is also added but is not fixed. In the absence of wormhole stabilization mechanisms, hence without the relation of $S_{\rm dS}$ and $\tau_0$, we would first like to explore the possible parameter space and identify interesting ones to focus.

\paragraph{Scaling relation.}
Our solution has a scaling relation among the first three of them. The two-point correlations are approximately invariant under the scaling by a constant $c$,
\beq
S_{\rm dS} \,\to\, S_{\rm dS} \cdot c, \quad \tau_0 \,\to\, \tau_0/c, \quad k_{\rm max}^c \,\to \, k_{\rm max}^c \cdot c.
\label{eq:scaling}\eeq
This becomes more exact in the small mass $m\ll 1$ limit. 
This is numerically demonstrated in \Fig{fig:scaling}, where $k_{\rm max}^c$ versus $\tau_0$ is shown for two fixed choices $S_{\rm dS}=76$ and $760$, which are related by $c=1/10$. The scaling can be traced back to the $Z_{\rm rat}$ \Eq{eq:phasecompet} in the phase transition mechanism, which is approximately calculated at the phase transition as 
\beq
\log Z_{\rm rat} \,\propto\, k_{\rm max}^c \log \frac{1}{\tau_0} \,\simeq\, S_{\rm dS},
\label{eq:Zrat0} \eeq
where the first approximation is numerically demonstrated in the left-bottom panel of \Fig{fig:obs1-1}; the average value of the mode-function ratio $\sqrt{{\rm Im}{\cal A}_k^{HH}/{\rm Im}{\cal A}_k^{WH}}$ is proportional to $\log 1/\tau_0$ in the small $\tau_0 \ll 1$ limit. As $1/\tau_0$ also sets the scale of $k_{\rm max}^c$ (hence, $k_{\rm max}^c \propto 1/\tau_0$), this proves the approximate scaling relation in \Eq{eq:scaling}.

\paragraph{Scrambling regime.}
We can divide the whole parameter space in \Fig{fig:scaling} into three regimes, separated by vertical lines. The division reflects different timescales of the phase transition, $k_{\rm max}^c$. Notably, in the majority of the parameter space, the fast scrambling timescale, $k_{\rm max}^c \sim S_{\rm dS}$ (or, $Ht \sim \log S_{\rm dS}$), turns out to be relevant to the bra-ket wormhole physics.

The fast scrambling timescale is when the information thrown into a horizon becomes thermalized enough to be able to leak out of the horizon~\cite{Hayden:2007cs,Maldacena:2015waa}. Why can this timescale be relevant to the two-point correlation across the horizon, via bra-ket wormholes? At first sight, this sounds unreasonable. Wormhole effects are topologically suppressed by $e^{-S_{\rm dS}}$, so that they are expected to be relevant only after a large multiplicity of $e^{S_{\rm dS}}$ is involved~\cite{Page:1993wv,Arkani-Hamed:2007ryv}; in particular, distinguishing a mixed state from a pure state in the Hilbert space of dimension $e^S$ requires to measure $e^S$ states. This timescale is around the Page time $Ht \sim S_{\rm dS}$. Indeed, in the black hole physics, replica wormholes begin to dominate around this timescale to modify the fine-grained entropy of the black hole~\cite{Penington:2019kki,Almheiri:2020cfm}, completing the Page curve. But correlation functions can be different.

\begin{figure}[t]
    \centering
    \includegraphics[width=0.50\linewidth]{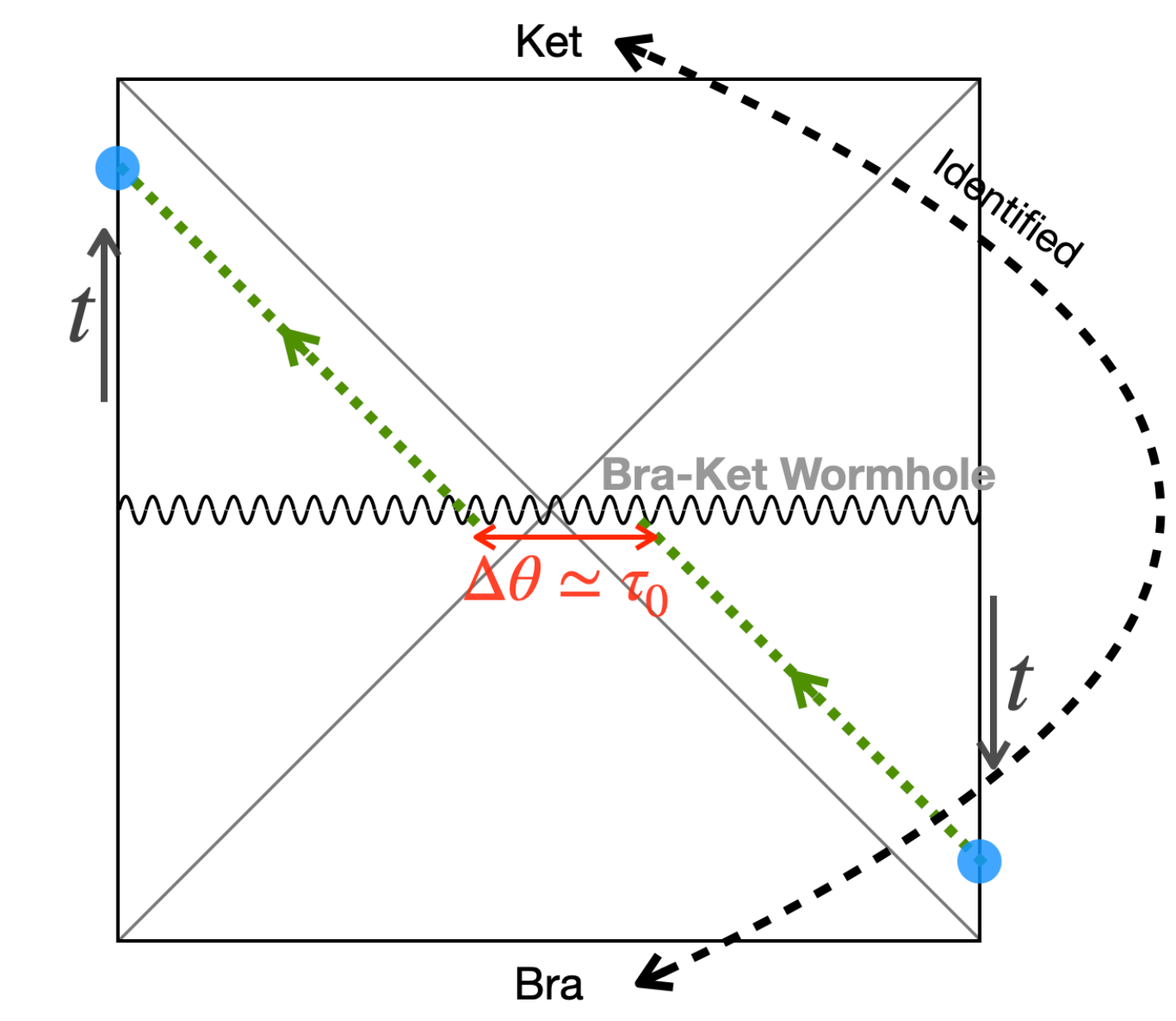}
    \caption{How the Euclidean shortcut via a bra-ket wormhole enhances the antipodal correlation by creating a shorter null geodesic (dotted). For given $\tau_0$, the minimum time evolution (modular flow) of $t \simeq \log (4/\tau_0)$ is needed to create the desired spatial shift $\Delta \theta \simeq \tau_0$. This is effectively the same as the shock-wave protocol to create a null geodesic across the horizon~\cite{Shenker:2013pqa,Gao:2016bin,Maldacena:2017axo,Aalsma:2021kle}, hence related to the scrambling physics.}
    \label{fig:scramble}
\end{figure}

One interpretation of the relevance of the scrambling time involves the shortcut via a Euclidean wormhole shown in the Penrose diagram of \Fig{fig:scramble}. This is a bra-ket wormhole, whose Euclidean region is denoted by the central curly line; we treat our effective wormhole after tracing in this way. When the time evolution (modular flow so that the bra and ket sides are located on the same time values) of two antipodal points is small, no null geodesics can connect them, and the correlation is suppressed. There exists the minimum amount of modular flow $t$ such that the given $\tau_0$ can geometrically allow a new null (hence shorter) geodesic between them, enhancing the correlation. At this point, our wormhole metric allows
\beq
Ht \,\simeq\, {\rm arccosh}(1/\sin (\tau_0/2) )\,\simeq\, \log (4/\tau_0), \qquad \Delta \theta \,\simeq\, \tau_0.
\label{eq:scramble-t}\eeq
This picture of the Euclidean shortcut via a wormhole is effectively the same as the shock-wave protocol to create a null geodesic across the horizon~\cite{Shenker:2013pqa,Gao:2016bin,Maldacena:2017axo,Aalsma:2021kle}. A signal sent from the South Pole in the past grows chaotically to ${\cal O}(1)$, and finally modifies the geometry to create a shorter null geodesic across the horizon. This growth is exponential near the horizon, yielding the scrambling timescale. Our estimate in \Eq{eq:scramble-t} becomes the scrambling time $Ht\sim \log S_{\rm dS}$ if we identify $1/\tau_0 \sim S_{\rm dS}$. And this is the relation that is observed numerically in the scrambling regime in \Fig{fig:scaling}.

In Ref.~\cite{Jung:2025rip}, authors argued the same result---the scrambling timescale of the phase transition---and confirmed it in random state models. There it was argued based on the competition between the topological suppression $e^{-S}$ and the bra-ket wormhole action $e^{\alpha L}$ for some $\alpha>0$ (due to the independence of the local bra-ket wormhole geometry on the dS boundary length $L$). Interpreting $L \sim e^{Ht}$, they obtained the scrambling time as the phase transition timescale. In our model, a similar competition is responsible for the scrambling timescale. The mode counting in \Eq{eq:phasecompet}, $Z_{\rm rat} = \prod^{k_{\rm max}}_k \sqrt{\frac{{\rm Im} {\cal A}_k^{HH}}{{\rm Im} {\cal A}_k^{WH}}} \,\sim\, e^{\alpha' k_{\rm max}}$ for some $\alpha'>0$ (see \Eq{eq:Zrat0} and \Fig{fig:obs1-1}), is actually \emph{multiplicative}  (rather than additive), so that the transition can occur exponentially earlier than the Page time. The reason for the necessary positivity of $\alpha>0$ is the negative Casimir energy in dS+CFT~\cite{Maldacena:2019cbz,Jung:2025rip}, while it is the relaxed KSW or the enhancement of low-$k$ correlations in our solution in dS+QFT.

The fast scrambling timescale is also parametrically close to the timescale estimated to be the lifetime of dS by Trans-Planckian censorship conjecture (TCC) and swampland~\cite{Bedroya:2019snp}, $k_{\rm max}^c \sim \sqrt{S_{\rm dS}}$ (or, $Ht \sim \tfrac{1}{2}\log S_{\rm dS}$). The dS phase may cease to continue there or some non-trivial effects may show up to modify this conclusion.

In the rest of the paper, we will focus on the scrambling regime by choosing the following benchmark parameters (marked as a cross in \Fig{fig:scaling}):
\beq
S_{\rm dS} \,=\,76, \quad \tau_0 \,=\, 0.01, \quad m \,=\, 0.1, \quad k_{\rm max} \,=\, 150.
\label{eq:benchmark} \eeq
The wormhole stabilization, if available in the future, will only change $\tau_0$ for the given $S_{\rm dS}$. 
The results of the Page regime will be qualitatively similar, except for the possibility of eternal inflation which may require another non-trivial modification. Another reason is to avoid huge computation time for exponentially long path integral for the Page regime.

Lastly, the leftmost region in \Fig{fig:scaling} is probably invalid. Here, the phase transition occurs too much earlier than the scrambling time. This is probably inconsistent with causality. Thus, we expect that a true wormhole stabilization mechanism will avoid this region.

The scalar mass is less of concern. Since $\phi$ may represent the slow-roll inflaton, this would rather be light. Too heavy ones also spoil the scaling behavior so that the analysis will be complicated. But too light ones, on the other hand, may exhibit remnants of infrared(IR) divergence that arise at $m=0$. Two origins of IR divergences. One is the massless Green's function, which affects both HH and WH. The other is maximal mixing of $\phi_1$ and $\phi_2$ in two separate universes ($b_k = a_k$ in \Eq{eq:braketS}), which leads to divergence after tracing out one of the universes. This further enhances IR divergence of WH contributions.  Thus, we take $m=0.1$ in the benchmark.

\subsection{Observable 1 - Comparing longer and shorter inflation} \label{sec:obs1}

\begin{figure}[t]
    \centering
\includegraphics[width=0.47\linewidth]{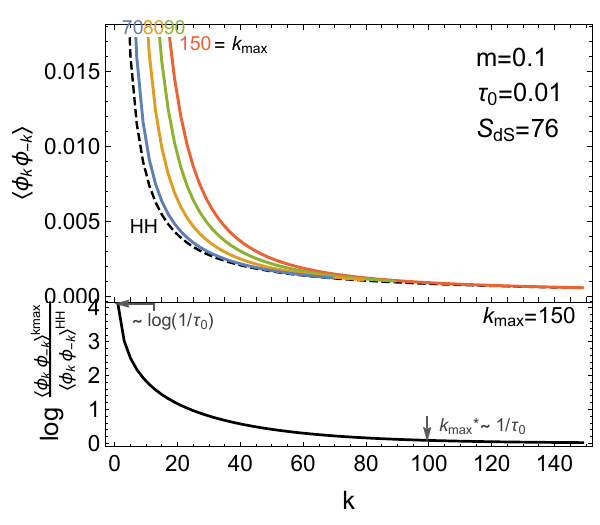}
    \includegraphics[width=0.52\linewidth]{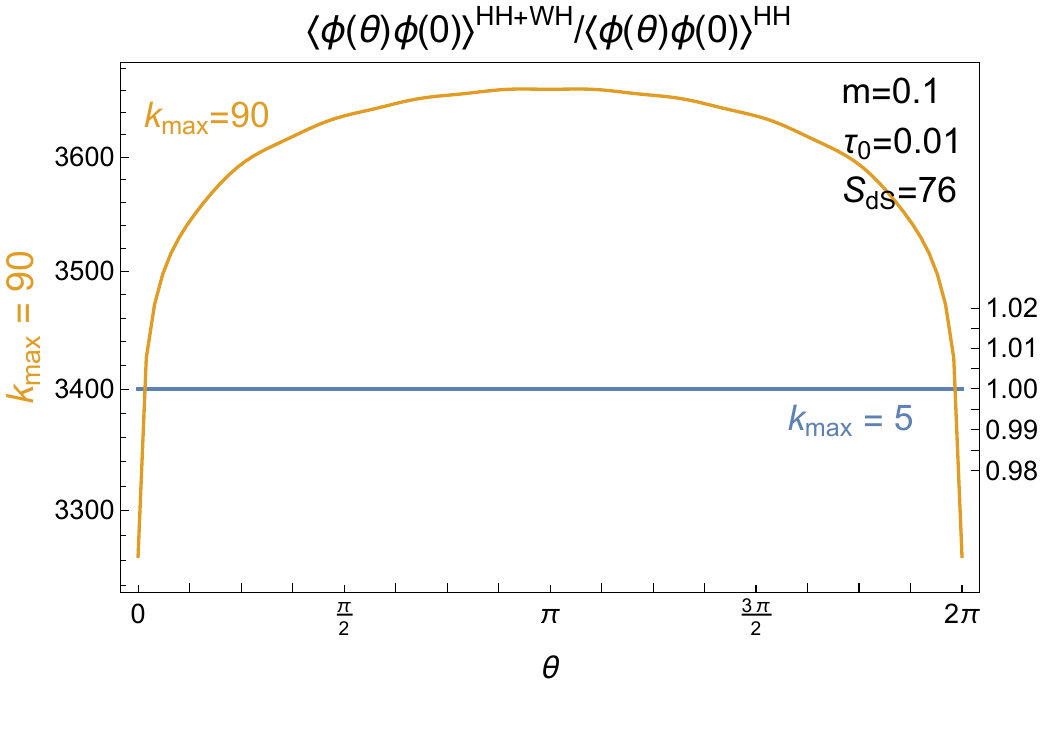}
    \caption{Observable 1: Basic features of wormhole-dominated correlations. $\langle \phi_k \phi_{-k} \rangle$ (left) and $\langle \phi(\theta) \phi(0) \rangle$ (right), compared between different $k_{\rm max}$-universes. Wormhole contributions are visible as deviation from HH results (dashed) for $k_{\rm max} \geq k_{\rm max}^c \simeq 60$. Small-$k$ or long-range correlations are enhanced with maximum between antipodal points. (left-bottom): the range of each mode-function ratio is set by $\tau_0$ for given $S_{\rm dS}$.}
    \label{fig:obs1-1}
\end{figure}

Observable 1 is to compare dS prepared states of different universes with different $k_{\rm max}$, including all modes up to $k_{\rm max}$. Two-point correlations in the full state are calculated by \Eq{eq:obs1} and (\ref{eq:spcorr1}). Questions for Observable 1: Given the gravitational scale $S_{\rm dS}$, is phase transition and its timescale consistent with information physics and the chaotic nature of a horizon? Here, we attempt to interpret $k_{\rm max}$ as some kind of time internal to dS, with aforementioned caution.

Above all, \Fig{fig:obs1-1} demonstrates the phase transition mechanism. Wormhole effects on $\langle \phi_k \phi_{-k} \rangle = {\rm Tr}[ \rho \,\phi_k \phi_{-k} ]$ are visible (as deviations from HH expected results) only for universes with large enough $k_{\rm max} \geq k_{\rm max}^c \sim 60$. Once the state $\rho$ is dominated by wormhole effects with large $k_{\rm max}$, the deviation is visible predominantly through enhancements of small-$k$ correlations. These are as expected by the mechanism in \Sec{sec:mech}. 
The fact that wormholes dominate only after long duration of dS stage may be consistent with its possible role of resolving information problem that appears only late.

\Fig{fig:obs1-1} also makes a consistency check that correlations of large-$k$ modes always agree with leading-order HH results. 
It is not that the HH term (first term) in \Eq{eq:obs1} dominates the wormhole(WH) term (second term) for large-$k$ correlations, but that ${\rm Im}A_k^{WH} \simeq {\rm Im}A_k^{HH}$ for large $k>k^*\simeq 100 (\sim 1/\tau_0)$ in \Eq{eq:obs1-2}, so that even for WH-dominated states large-$k$ correlation is always that of HH result. 

In the left-bottom panel, the log of each mode-function ratio at $k_{\rm max} = 150$ is plotted, showing that the area under the curve (essentially the $Z_{\rm rat}$) scales with $\log Z_{\rm rat} \propto \frac{1}{\tau_0} \log \frac{1}{\tau_0}$. As $1/\tau_0$ sets the scale of $k_{\rm max}^c$ and $k_{\rm max}^*$, this proves the scaling relation in \Eq{eq:scaling} and the appearance of the scrambling timescale in \Fig{fig:scaling}. 

\Fig{fig:obs1-1} right panel shows real-space correlations. 
Unlike scale invariant HH results, WH-dominated spatial correlation is largest between antipodal points $\theta = \pi$, while rapidly decreasing for short-distance correlation $\theta \to 0,2\pi$ (but still larger than HH results). This enhancement of long-range correlation was expected if WH solution is to be consistent with KSW. And we also argued that this might be attributed to wormhole's stereotypical property of entangling across the horizon, shortening a geodesic.

\begin{figure}[t]
    \centering
    \includegraphics[width=0.49\linewidth]{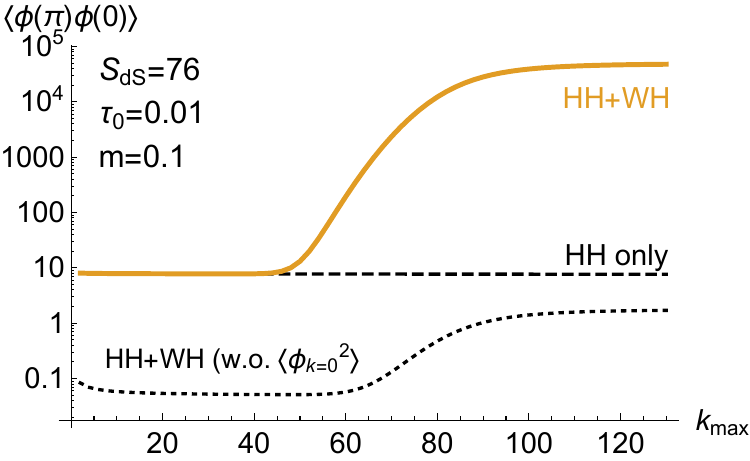}
    \includegraphics[width=0.49\linewidth]{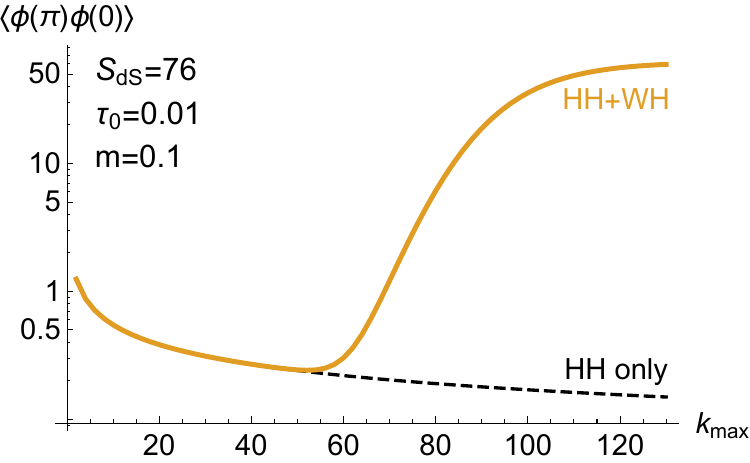}
    \caption{Observable 1: Correlation does not decay forever, but ramps up and reaches a plateau, due to bra-ket wormhole contributions. This feature may be consistent with the finiteness of dS horizon entropy. Quasinormal decays of superhorizon modes apparent in the heavy mass case (right panel) may reflect the chaotic nature of dS. Excluding zero modes ($k=0$) results in the overall decrease of correlation, but main features remain.}
    \label{fig:obs1-3}
\end{figure}

\Fig{fig:obs1-3} shows an important result: correlation does not decay indefinitely with time $k_{\rm max}$, but ramps up and plateaus at late time. Similar behavior was found in the AdS-black hole system~\cite{Saad:2018bqo,Saad:2019lba,Saad:2019pqd}, which was largely expected based on the finiteness of Hilbert space~\cite{Maldacena:2001kr,Witten:2001kn,Goheer:2002vf}. It was expected also in dS~\cite{Aalsma:2022eru,Chapman:2022mqd,Mirbabayi:2023vgl}. Thus, we conclude that this decay-ramp-plateau behavior is an important indication that bra-ket wormholes may really be relevant to the finiteness of dS horizon entropy.

As a bonus, \Fig{fig:obs1-3} also exhibits quasinormal decays of superhorizon modes, shown as the slow decrease of HH result, more apparently in the heavy mass case ($m=0.49$). Technically, this effect is incorporated by $(2\pi/\ell)^{-\Delta_-} = k_{\rm max}^{-\Delta_-}$ factor in \Eq{eq:Sqft1}.  This exponential decay is originated from the exponential expansion of dS universe, but it also suggests that dS is a chaotic quantum system~\cite{Shenker:2013pqa,Maldacena:2015waa,Aalsma:2020aib}.

One concern is about the zero mode ($k=0$) contribution. Formally, there is IR divergence in $m=0$, and large enhancement for finite $m$. Excluding zero modes in \Fig{fig:obs1-3} significantly reduces the overall correlation, but main features of phase transition, its timing, and decay-ramp-plateau all remain. Zero modes affect only real-space correlations of Observable 1. Without clear resolution of how to treat them, we are content with keeping zero modes everywhere else.

\subsection{Observable 2 - CMB-like correlation after reheating} \label{sec:obs2}

Observable 2 is CMB-like two-point correlations measured by an observer in the flat (or FRW) space after reheating. At each time of measurement (parameterized by $k_{\rm min}$), we combine only $k_{\rm min} \leq k \leq k_{\rm max}$ modes to calculate real-space correlations in \Eq{eq:spcorr2}.
$k_{\rm min}$ serves as a time after reheating; decreasing $k_{\rm min}$ means time running forward. $k_{\rm max}$ is fixed, and the dS prepared state is fixed. This observable is what a post-reheating observer in the given universe would really measure as a function of his/her cosmological time, via CMB or large scale structures.

Two questions for Observable 2. Is CMB-like observation of dS prepared state consistent with complementarity and resolution of information problem? What are observational features of wormhole dominance, which our cosmological descendants may be able to use to test this idea and understand fundamental physics of dS?

\begin{figure}[t]
    \centering
    \includegraphics[width=0.49\linewidth]{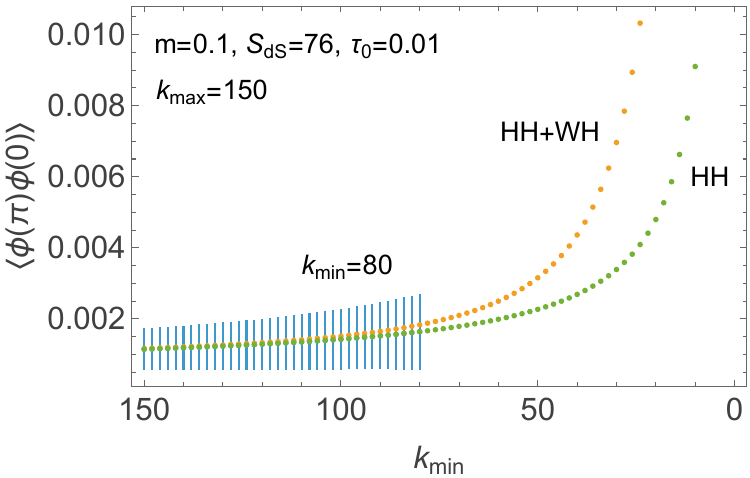}
    \includegraphics[width=0.49\linewidth]{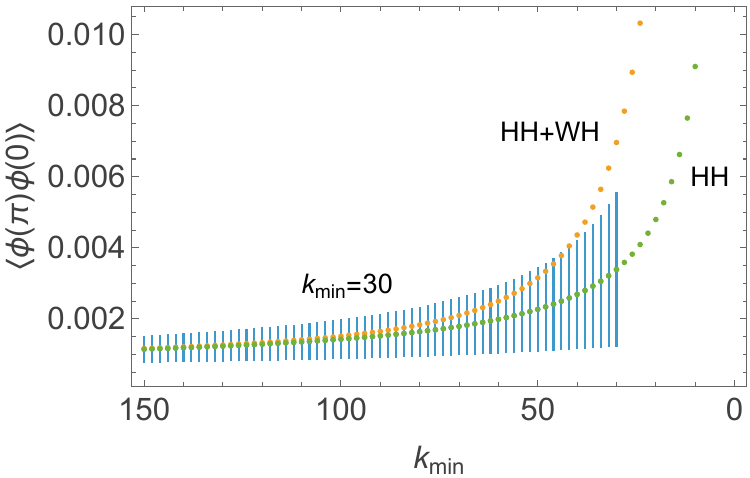}
    \includegraphics[width=0.49\linewidth]{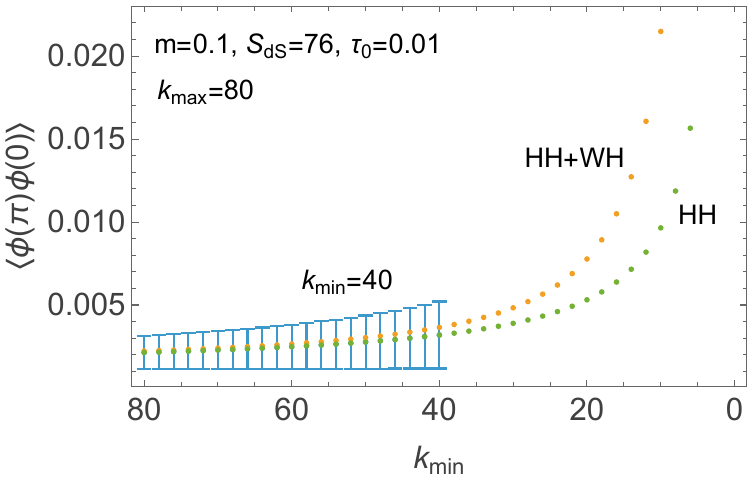}
    \includegraphics[width=0.49\linewidth]{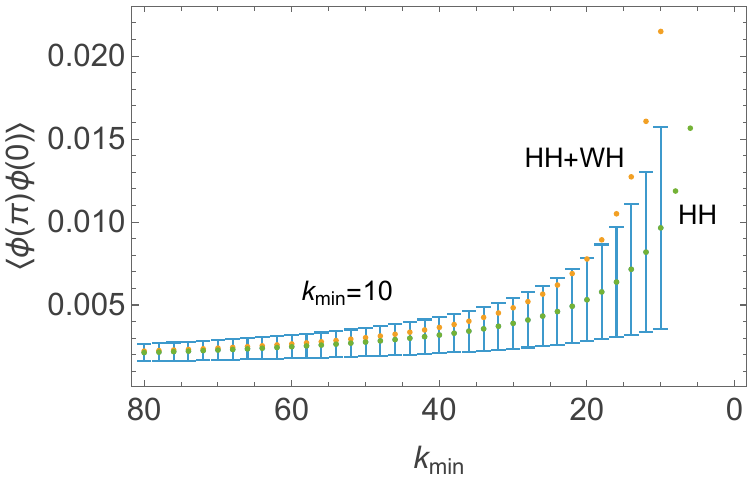}
    \caption{Observable 2: HH+WH as functions of time after reheating $k_{\rm min}$ (decreasing from $k_{\rm max}$) are compared with HH alone. Overlaid are irreducible $1\sigma$ error bars from \emph{cosmic variance} of subhorizon modes at the marked time $k_{\rm min}$. Upper two panels with $( k_{\rm max}, k_{\rm min} )= (150,80),\, (150,30)$, and lower two with $(80,40),\, (80,10)$; the upper and lower panels are distinguished by $k_{\rm max}$ versus $k^* \simeq 100$. In both cases, early CMB observers would not be able to distinguish wormhole effects.}
    \label{fig:obs2}
\end{figure}

The key result is shown in \Fig{fig:obs2} that CMB initially looks indistinguishable from HH predictions (Bunch-Davies), but it gradually deviates at late times due to wormhole effects.

How phase transition occurs while being chronologically consistent is notable, different from that in Observable 1. Consider first the case with $k_{\rm max} > k^* (\simeq 100)$ (the upper panels of \Fig{fig:obs2}). At early times, only a small number of largest-$k$ modes are observed. They by themselves resemble HH, which was a consistency property. It is not that these largest $k$-modes are generated during the HH dominance, rather they are the ones generated at the last stage of dS where wormholes dominate most. But they resemble HH by themselves, even though the fixed state, a sum of the whole modes, is dominated by wormhole effects. As time goes on ($k_{\rm min}$ decreasing), smaller $k$-modes re-enter, and discernible wormhole contributions gradually appear~\cite{Page:1993wv}.

Consider now the case with $k_{\rm max} < k^*$ (the lower panels of \Fig{fig:obs2}). In this case, the early CMB modes in principle are different from HH results. But \emph{cosmic variance} fundamentally limits the early-time distinction of WH+HH from HH results. This is well illustrated in \Fig{fig:obs2} by $1\sigma$ error bars of cosmic variances. For a simple proxy of the 2D-universe cosmic variance, we take the 4D-universe (2D CMB sky) result for the relative error: $\Delta C_k / C_k \simeq \sqrt{2/(4k/k_{\rm min}+1)}$, where it was used that the spherical harmonics $\ell$ for the $k$-mode when the $k_{\rm min}$-mode just re-enters the horizon is approximately $\ell(k) \simeq 2k/k_{\rm min}$. Thus, again  CMB observers cannot distinguish wormhole effects until late. It is interesting that one form of fundamental constraint---cosmic variance---makes observations consistent with another fundamental physics---CMB complementarity.

Why is this chronological order of CMB observation consistent with information content of the observer?
Shortly after reheating, CMB observers just see a small amount of information that match with the HH vacuum, never able to disentangle exponentially suppressed higher topology effects. As the state would essentially look thermal, the entropy grows. Only after wormhole effects are distinguished at late times, the entropy decreases. This order is consistent with the aforementioned discussion of pure versus mixed-state distinction.
This chronological phenomenon is also analogous to that of black hole Hawking radiation and the recovery of information(unitarity) only after the Page time. 

In a slow-roll inflation, the dS Page time is when the dS horizon entropy is saturated by superhorizon modes generated during inflation. 
Longer inflation essentially becomes an eternal inflation. 
What happens to the CMB observation? It was argued that density fluctuations generated during the eternal regime are ${\cal O}(1)$ so that no CMB observations after the dS Page time can be made, not violating complementarity~\cite{Bousso:2006ge,Arkani-Hamed:2007ryv}. Our phase transition that may occur much earlier than the Page time (in Observable 1) may cure or change related physics and observations well before the saturation; but the possibility of Page-time phase transitions are still open in our solution.

A few more minor thoughts on CMB complementarity. One missing ingredient in our calculation might be that $S_{\rm dS}$ in a slow-roll becomes maximal toward the end of inflation, making wormhole effects relatively smaller there. And these last modes are the ones that will be observed at early times by CMB observers. It is also an interesting question by itself how to implement slow-rolling in gravitational path integrals, or more generally the response of the gravity action to the matter sector which is absent in the JT model. One can also make an anthropic argument. No advanced observers can exist right after reheating; it always takes some time for universes to evolve to accommodate advanced observers. Notably, this anthropic argument is unique to dS-flat setup, as gravitating and flat regions are timelike separated; black hole observers can exist simultaneously with the black hole.

\begin{figure}[t]
    \centering
    \subfigure{\includegraphics[width=0.4\textwidth]{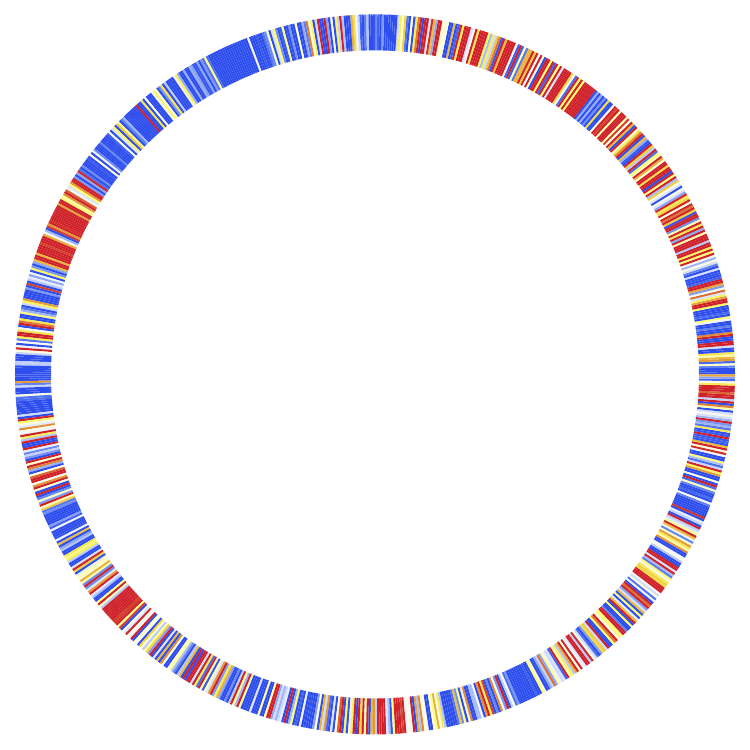}}
    \label{fig:tCMBHH}
    \subfigure{\includegraphics[width=0.4\textwidth]{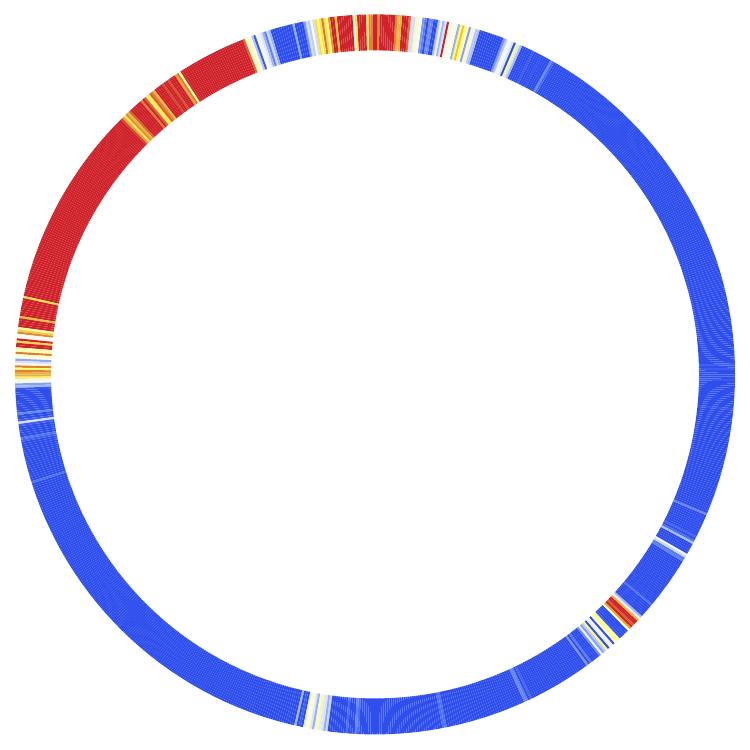}}
    \label{fig:tCMBWH}
    \caption{Example realization of the CMB field $\phi(\theta)$, without(left) and with(right) wormhole contributions. $k_{\rm max}=150, k_{\rm min}=0, m=0.1, \tau_0 = 0.01, S_{\rm dS}=76$. Enhanced long-range correlations are clear.}\label{fig:toyCMBlight}
\end{figure}
\begin{figure}[t]
    \centering
    \subfigure{\includegraphics[width=0.4\textwidth]{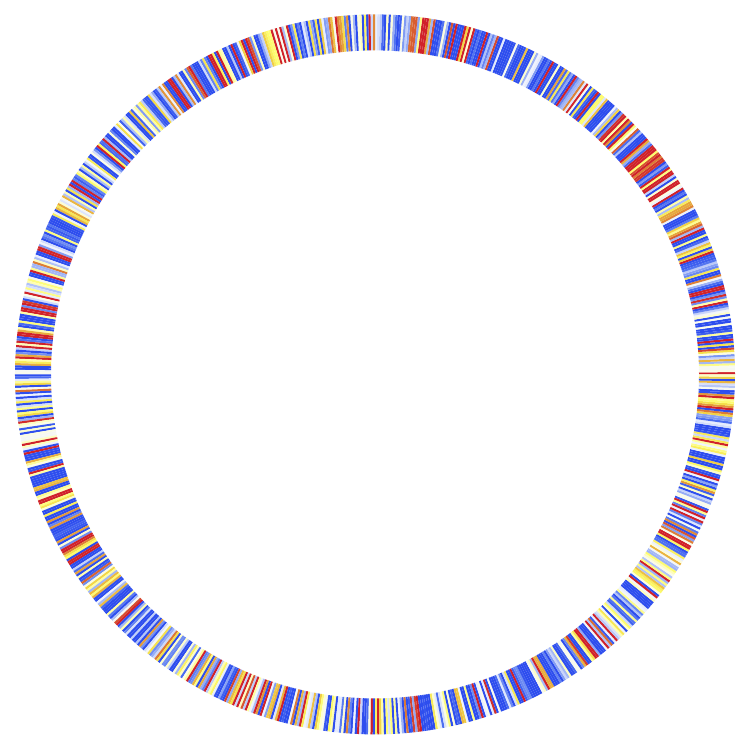}}
    \label{fig:tCMBHH01}
    \subfigure{\includegraphics[width=0.4\textwidth]{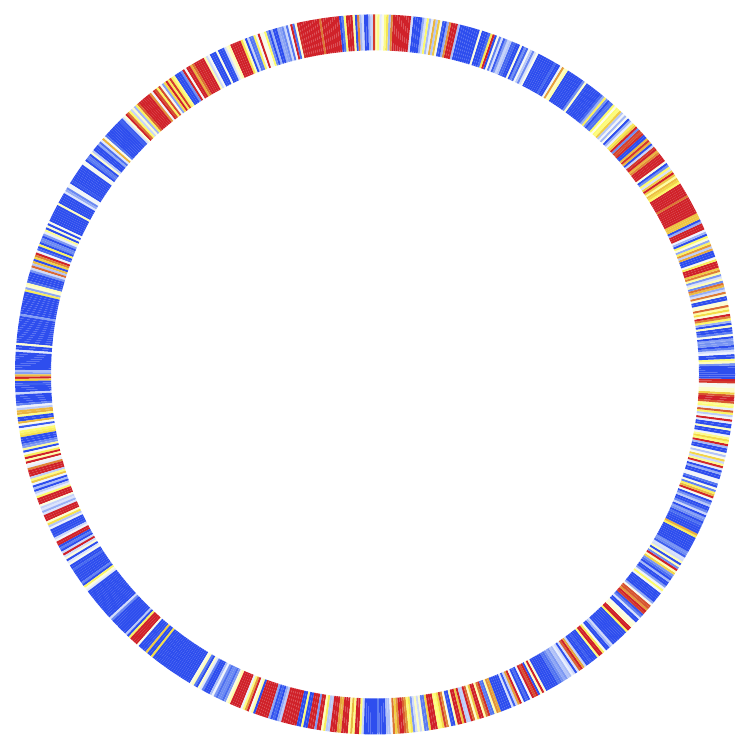}}
    \label{fig:tCMBWH01}
    \caption{Same as \Fig{fig:toyCMBlight}, but for heavier $m=0.49$. Enhancement of long-range correlation is less prominent.}
    \label{fig:toyCMBheavy}
\end{figure}

Lastly, \Fig{fig:toyCMBlight} and \ref{fig:toyCMBheavy} show example late-time CMB distributions on the 1-dimensional space; full results on the right panel is compared with results without wormholes on the left panel. In these sample realizations, field values $\phi_k$ are random selected according to the Gaussian probability with the width given by $\langle \phi_k \phi_{-k} \rangle$. 
The wormhole effects show up as stronger long-range correlations as expected. It is more prominent for light scalars. The breaking of scale invariance can also be measured as enhanced dipole anisotropies.

\section{Summary and Discussion} \label{sec:discussion}

In this work, we have studied whether bra-ket wormhole saddle solutions to two-point correlators exist and affect in a way (at least qualitatively) consistent with dS information problem, in the two-dimensional JT dS plus scalar QFT.
Above all, in \Sec{sec:WH}, we have found a bra-ket wormhole solution at NLO, in addition to a bra-bra and a ket-ket wormholes. To that end, we had to complexify the scale factor, which still satisfies the classicality and relaxed KSW conditions. 

\begin{figure}[t]
    \centering
    \includegraphics[width=0.5\linewidth]{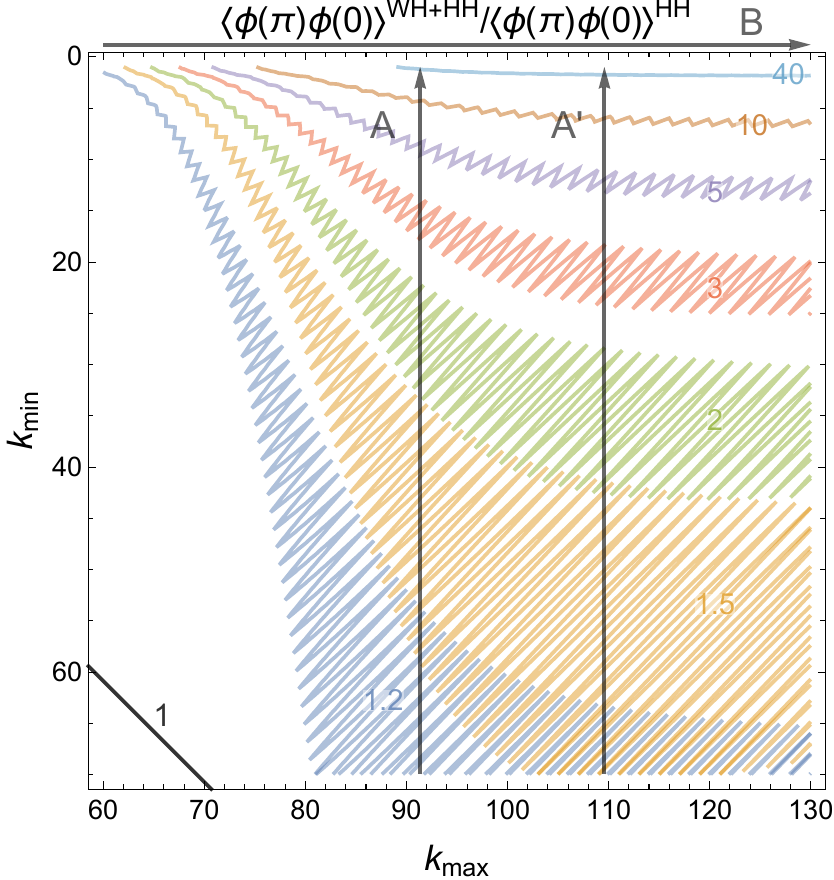}
    \caption{Two observables in the full two-dimensional domain \{$k_{\rm max}$, $k_{\rm min}$\}. Observable 1, comparing different universes with different $k_{\rm max}$, is along B. Observable 2, measuring CMB of a given universe with time after reheating $k_{\rm min}$, is along A. Also shown are highly oscillating contours of relative wormhole effects, telling its time-dependence in observables. $S_{\rm dS}=76$ and $\tau_0=0.01$.}
    \label{fig:obsall}
\end{figure}

In \Sec{sec:Obs-BKWH}, we have introduced two correlation observables of the dS prepared state by utilizing two time variables. The parameter spaces they are probing are summarized in the full parameter space $\{ k_{\rm max} , k_{\rm min} \}$ in \Fig{fig:obsall}. The Observable 1 (along $B$) compares dS prepared states of different universes, allowing to diagnose the dynamics of phase transition into wormhole dominance. The Observable 2 (along $A$ or $A'$ depending on which $k_{\rm max}$-universe) is basically the CMB observable as a function of observer's cosmological time after reheating. Each correlation observable was qualitatively consistent by itself, finding the decay-ramp-plateau--type behavior, phase transitions at about the scrambling time, and consistent chronological orders. These were possible due to (1) inflationary exit and re-enter, (2) small-$k$ enhancement of correlations, (3) phase transition mechanism via competition between mode counting and topological suppression, and (4) irreducible errors from cosmic variance.

A notable result was the possible relevance of scrambling physics to the correlation functions via bra-ket wormholes. We attempted to associate the shock-wave scrambling physics with the shortcut in our bra-ket wormhole geometry. We demonstrated the appearance of the scrambling timescale from the competition between \emph{multiplicative} mode counting and topological suppression. But Page-time phase transitions are still possible in a small range of the parameter space.
The concrete answer and interpretation in terms of entropy and chaotic aspects of dS requires to have a complete wormhole stabilization, which will provide the definite relation of $\tau_0$ and $S_{\rm dS}$. If scrambling physics turns out to be correct, our work suggests $1/\tau_0 \sim S_{\rm dS}$.

%

\paragraph{Implications.}
Our explicit wormhole solution may shed some light on dS-CFT correspondence. The expected asymptotic behavior of the scalar field in \Eq{eq:Phi_asymp}
\begin{equation}
    \Phi \,\sim\, \phi \,e^{-\Delta_- t} + \mathcal O\,e^{-\Delta_+ t} \,\simeq\, \phi\, (-\eta)^{\Delta_-} + {\cal O}\,(-\eta)^{\Delta_+}
\end{equation}
is confirmed in our explicit solution of wormhole too. Written in terms of the conformal time $\eta = -e^{-t}$, this makes it clear that the conformal invariance of 2d-scalar $\Phi$ implies 1d-CFT transformation of $\phi$ and ${\cal O}$ with conformal dimension $\Delta_-$ and $\Delta_+$, respectively. The resulting CFT correlation $\langle \phi(x) \phi(x') \rangle \sim |x-x'|^{-2\Delta_-}$ is reproduced exactly by the HH wavefunction, which is a well-known support of the conjecture of dS-CFT correspondence~\cite{Strominger:2001pn,Maldacena:2002vr}. In our wormhole solution, this correlation was modified even though the asymptotic form and the conformal dimensions $\Delta_\pm$ remain unmodified. It suggests that the corresponding CFTs may be modified by interactions between same kind of operators. This is a reminiscent of how traversable wormholes in AdS is generated by double-trace deformation that couples the two boundary CFTs~\cite{Gao:2016bin,Maldacena:2017axo,Maldacena:2018lmt} preparing the entangled thermofield double state~\cite{Cottrell:2018ash}. We leave detailed check of correspondence in this solution for future work.

Another implication of modified two-point functions is on the stochastic description of inflation~\cite{Starobinsky:1986fx,Starobinsky:1994bd}. This description is based on the local interplay of deterministic classical rolling (drift) of inflaton($\Phi$) and random fluctuation (diffusion) from two-point function. Instead of scale invariant diffusion $\sim H/2\pi$ from the HH two-point function, our wormhole solution will lead to scale-dependent diffusion, via enhanced long-wavelength correlation at early stage of slow-rolling. This may modify the global probability distribution of the inflaton field value, potentially affecting `natural' outcomes of inflaton-dependent `inputs' to the big-bang universe~\cite{Giudice:2021viw,Jung:2021cps}.

\acknowledgments
We thank Hyung Do Kim, Seok Kim, Jiwoo Park, Dong-han Yeom, Junggi Yoon for valuable comments and encouragement. We are supported by Grant Korea NRF2019R1C1C1010050 and RS-2024-00342093.

\appendix

\section{Scalar field solutions and correlators}\label{app:field_sol}

\subsection{Hartle-Hawking} \label{app:HH}

We first analyse the scalar field on the Hartle-Hawking geometry with metric
\begin{equation}
    ds^2 = -dt^2 + \cosh^2t\,d\theta^2.
\end{equation}
The governing equation for  the scalar field $\Phi_k(t)$ of mass $m$ is written by
\begin{equation}
    \partial_{t}^{2} \Phi_k(t)+\tanh t\, \partial_{t} \Phi_k(t)+ \left(\frac{k^2}{\cosh ^{2} t}  +m^{2}\right) \Phi_k(t)\,=\,0.
\end{equation}
We impose boundary conditions at a late-time cutoff surface,
\begin{equation}
    \Phi_k(t_\epsilon) = \phi(k),
\end{equation}
with $t_\epsilon = \log(2/\epsilon)$. The solution is required to be regular on the full geometry, including the Euclidean cap. Regularity at the pole of the Euclidean hemisphere removes the negative-frequency modes and prepares the Hartle--Hawking state, which coincides with the Bunch--Davies vacuum in the Lorentzian regime.

The resulting solution can be written as
\begin{align}
    \Phi_k(t) \,=\, \phi(k) \frac{u_k(t)}{u_k(t_\epsilon)},~~u_k(t) \,=\,\frac{i^{\Delta_-} \pi }{2^{\nu}\Gamma(\nu)}\left( \tanh^2 t-1\right)^{1/4} \left( P_{k-1/2}^\nu(\tanh t) - \frac{2 i}{\pi} Q_{k-1/2}^\nu(\tanh t) \right),
\end{align}
where $u_k(t)$ is the linear combination of associated Legendre functions that is regular on the Euclidean cap. For \(t_\epsilon \gg 1\), its asymptotic behavior is
\begin{align}
   u_k(t_\epsilon) &\,=\, \left[\epsilon^{\Delta_-}+\,i^{2\nu}\,\frac{\Gamma(-\nu)}{4^\nu\Gamma(\nu)}\,\frac{\Gamma(\Delta_+ + k)}{\Gamma(\Delta_- +k)}\,\epsilon^{\Delta_+}\right].
\end{align}
In terms of the boundary value $\phi$, the on-shell action is
\begin{align}\label{eq:HHaction_app}
    iS[g_{\rm cl},\Phi_{\rm cl}] &\,=\, \frac i2 \int d\theta\,a(t)\Phi_{\text{cl}} (t,\theta) \partial_{t} \Phi_{\text{cl}} (t,\theta)  \bigg|_{t=t_\epsilon}\\\nonumber 
    &\,=\, i\pi \sum_k \phi(k) \phi(-k) \frac{1}{\epsilon} \frac{\partial_t u(t)}{u(t_\epsilon)}\Bigg|_{t=t_\epsilon} \\\nonumber
    &\,=\, -i\pi \sum_k \phi(k) \phi(-k) \frac{1}{\epsilon}\left[ \Delta_- + 2\nu\, i^{2\nu}\,\frac{\Gamma(-\nu)}{4^\nu\Gamma(\nu)}\,\frac{\Gamma(\Delta_+ + k)}{\Gamma(\Delta_- +k)} \epsilon^{2\nu} + \cdots  \right].
\end{align}
The first term in the square bracket does not contribute to the observables since it cancels upon taking the modulus squared. The imaginary part of the second term is the leading one in the computation with the factor, $\epsilon^{-1+2\nu} = \left( \frac{2\pi}{\ell}\right)^{-1+2\nu}$.  This reproduces the standard Bunch--Davies two-point function.

The density matrix for the boundary configurations is then
\begin{equation}
    \rho(\phi) \,\propto\, \big|e^{iS}\big|^2 \,=\, \exp \left[-\sum_k \left( \frac{2\pi}{\ell}\right)^{-2\Delta_-}\frac{4\pi^2}{4^\nu \Gamma[\nu]^2} \frac{\Gamma(\Delta_+ + k)}{\Gamma(\Delta_- +k)} \phi(k)\phi(-k)\right].
\end{equation}
The two-point function follows from a Gaussian integral over the boundary fields,
\begin{equation}
    \int D\phi \,\phi(k)\phi(-k) \rho(\phi) \,=\,\left( \frac{2\pi}{\ell}\right)^{2\Delta_-}\frac{4^\nu \Gamma[\nu]^2} {4\pi^2}\frac{\Gamma(\Delta_- +k)} {\Gamma(\Delta_+ + k)},
\end{equation}
which matches the expected Bunch--Davies correlator, reproducing the results in \cite{Maldacena:2019cbz} in the global coordinate.

In the gravity side, the bulk gravitational action reduces to a boundary term, parametrized by the Schwarzian modes $\theta$~\cite{Maldacena:2019cbz,Cotler:2019nbi}. 
\begin{equation}\label{eq:HHgravact}
    S_{\rm grav} 
      \,=\, \frac{\varphi_0}{4G_2}\,\chi + \frac{\varphi_r}{8\pi G_2}\int_0^{2\pi}d\theta\,\left(\{f(\theta),\theta\}+\frac{f'(\theta)^2}{2}\right),
\end{equation}
where \(\chi=-i\) for Hartle-Hawking and \(f(\theta)\) parametrizes the boundary curve. The Schwarzian derivative is defined as
\begin{equation}
    \{f(x),x\} \,\equiv\, \frac{f'''(x)}{f'(x)}-\frac{3}{2}\frac{f''(x)^2}{f'(x)^2}.
\end{equation}
The topological term in Eq.~\eqref{eq:HHgravact} contributes a factor \(\exp(S_{dS})\), making the Hartle–Hawking contribution dominant over higher-genus saddles, while the Schwarzian term vanishes upon setting \(f(\theta)=\theta\) before squaring the wave function to extract probabilities.

\subsection{Two-boundary wormhole} \label{app:WH}

To capture the NLO contributions from wormhole saddles, we now solve the scalar equation on a different background. A wormhole scale factor is
\begin{equation}
    a(t) \,=\, \frac12 \left( e^t +e ^{i\tau_0 -t}\right).
\end{equation}
The EoM for the scalar field on this geometry becomes 
\begin{equation}
    \partial_{t}^{2} \Phi_k(t)+\tanh (t-i\tau_0/2)\, \partial_{t} \Phi_k(t)+ \left(\frac{k^2}{e^{i\tau_0}\cosh ^{2} (t-i\tau_0/2)}  +m^{2}\right) \Phi_k(t)\,=\,0.
\end{equation}
A general solution can be written as
\begin{align}
   \upsilon_k(t) \,=\, e^{t\left(\frac{1}{2}-\nu\right)+i\nu\tau_0}&\left(e^{2t}+e^{i\tau_0}\right)^{-\tilde{k}} \Biggl[C_2\, {}_2F_1\Bigl(\frac{1}{2}-\tilde{k},\frac{1}{2}-\tilde{k}-\nu,1-\nu,-e^{2t-i\tau_0}\Bigr) \nonumber\\[1mm]
    &~~+ C_1\,e^{2\nu t+i(2\pi-\tau_0)\nu} \, {}_2F_1\Bigl(\frac{1}{2}-\tilde{k},\frac{1}{2}-\tilde{k}+\nu,1+\nu,-e^{2t-i\tau_0}\Bigr)\Biggr],
\end{align}
where we defined \(\tilde k = e^{-i\tau_0/2} k\).

We impose boundary conditions at the two asymptotic boundaries, located at $t = t_\epsilon,\,-t_\epsilon+i\tau_0$
\begin{equation}
    \Phi_k(t_\epsilon) \,=\, \phi_1(k), \qquad \Phi_k(-t_\epsilon + i\tau_0) \,=\, \phi_2(k).
\end{equation}
These conditions fix the coefficients $C_1,C_2$ as
\begin{align}
C_1 &\,=\, 
\,\frac{
e^{-i\pi\nu}\,2^{\Delta_-}\,
e^{-\frac{i}{2}\bigl(1 - 2 \tilde k\bigr)\tau_0}
\,
\Gamma\!\left(\tfrac{1}{2} - \tilde k + \nu\right)\,
\Gamma\!\left(\tfrac{1}{2} + \tilde k + \nu\right)
}{
\Gamma(\nu)\,\Gamma(1+\nu)
}\Bigl[\phi_1 
- \phi_2 \cos\!\bigl(\tilde k \pi\bigr)\,
\csc(\pi \nu)\Bigr],
\\
C_2 &\,=\, 
\,e^{i\pi\nu}\,2^{\Delta_-}\,
e^{-\frac{i}{2}\bigl(1 - 2\tilde k\bigr)\tau_0}
\phi_2 \,.
\end{align}
The field is constructed by the combination of
\begin{equation}
    \Phi_k (t) \,=\, \phi_1 (k) \frac{\upsilon_k(t)}{\upsilon_k(t_\epsilon)} \,=\, \phi_2 (k) \frac{\upsilon_k(t)}{\upsilon_k(-t_\epsilon+i\tau_0)}.
\end{equation}
Employing the asymptotic form of the $\upsilon(t) \simeq \phi \,a(t)^{-\Delta_-} + \mathcal O \,a(t)^{-\Delta_+}$, the boundary values of the fields and its time derivatives are
\begin{align}
    \Phi_k(t_\epsilon) &\,=\, \phi_1(k), \\
    \dot \Phi_k(t_\epsilon) &\,=\,- 2\nu \mathcal O_1(k) \epsilon^{2\nu}.
\end{align}
At the other boundary, the expressions are analogous with $1$ and $2$ interchanged. $\mathcal O_{1,2} = \tilde a\phi_{1,2}+\tilde b\phi_{2,1}$ can be obtained by expanding $u_k(t)$ near the boundary,
\begin{align}
\tilde a_k &\,=\, -\,\frac{
4^{\Delta_-}\,e^{i\nu\tau_0}\,\pi^2\,
\cos\!\bigl(\tilde k \pi\bigr)\,\csc(\pi\nu)
}{
\bigl(\cos(2 \tilde k \pi)+\cos(2\pi\nu)\bigr)\,
\Gamma\!\left(\tfrac12 - \tilde k - \nu\right)\,
\Gamma\!\left(\tfrac12 + \tilde k - \nu\right)\,
\Gamma(\nu)\,\Gamma(1+\nu)
},
\\
\tilde b_k &\,=\, \frac{
4^{\Delta_-}\,e^{i\nu\tau_0}\,\pi^2}{
\bigl(\cos(2 \tilde k \pi)+\cos(2\pi\nu)\bigr)\,
\Gamma\!\left(\tfrac12 - \tilde k - \nu\right)\,
\Gamma\!\left(\tfrac12 + \tilde k - \nu\right)\,
\Gamma(\nu)\,\Gamma(1+\nu)
}.
\end{align}
The on-shell action receives contributions from both boundaries,
\begin{align}
    S &\,=\, -2\nu\pi \sum_k \bigl( \phi_1(-k)\, \mathcal O_1(k) 
    + \phi_2(-k)\, \mathcal O_2(k) \bigr) \\
    &\,=\, -2\nu\pi \sum_k 
    \bigl[ \tilde a_k |\phi_1(k)|^2 
      + \tilde b_k [\phi_1(k) \phi_2(-k) + \text{c.c.}] 
      + \tilde a_k |\phi_2(k)|^2 \bigr].
\end{align}
As discussed above, the imaginary parts of the coefficients \(\tilde a_k\) and \(\tilde b_k\) determine the induced boundary correlators. The complex squaring of the path integral and the subsequent tracing-out procedure are carried out in \Sec{sec:trace_univ}, where we construct the physical density matrix relevant for observables.

\section{Stabilizing bra-ket wormhole by extra fermions} \label{app:Fermi-Stab}

To stabilize the wormhole, we can introduce the additional degrees of freedom for a fermion field whose action \cite{Loran:2004fu,Jiang:2020evx} is
\begin{equation}
    \mathcal{S} \,=\, -\frac12\int d^2x\, \sqrt{-g} \left( \tfrac i2\bar\psi\, \overleftrightarrow {\slashed{D}}\psi + m \bar\psi \psi \right) + \frac12 \int d\theta \sqrt{h} \,\bar\psi\gamma^0\psi.
\end{equation}
The covariant Dirac operator in ${\rm dS}_2$ is defined using the zweibein $e_a^{\mu}$ as $\slashed{D} = \gamma^\mu D_\mu = \gamma^a e^\mu_a \left( \partial_\mu + \frac{1}{4} \omega_\mu^{bc} \gamma_{bc} \right)$ where $\omega_{\mu}^{bc}$ is the spin connection and $\gamma_{bc} = \tfrac12 [\gamma_b,\gamma_c]$. The boundary term is included included in order to render the variational problem well-posed and guarantee $\delta S = 0$. The EoM for the two-component fermion field, $\psi = (u,v)^t$ is written as 
\begin{align}
    \left(\partial_t +\frac{\dot a(t)}{2a(t)}-\frac{i k}{a(t)}\right)u+mv \,=\, 0, \\
     \left(\partial_t +\frac{\dot a(t)}{2a(t)}+\frac{i k}{a(t)}\right)v+mu \,=\, 0,
\end{align}
working in the metric of $ds^2 = -dt^2 + a(t)^2 d\theta^2$. They reduce to the two coupled 2nd ODEs such as
\begin{align}
    &\partial_t^2 u(t,k) + \frac{\dot a(t)}{a(t)} \partial_t u(t,k) + \left[-m^2 + \frac{\ddot a(t)}{2a(t) } + \left(\frac{k}{a(t)}+i \frac{\dot a(t)}{2a(t)} \right)^2 \right]u(t,k) \,=\, 0, \\
    &v(t,k) \,=\, - \frac1m\left(\partial_t +\frac{\dot a(t)}{2a(t)}-\frac{i k}{a(t)}\right)u(t,k),
\end{align}
where $a(t)= e^{i\tau_0/2} \cosh(t-i\tfrac{\tau_0}{2})$ for the wormhole geometry we find in \Sec{sec:naiveWH}. After solving the EoM, one can expand the field near the asymptotic boundaries as done in the scalar field
\begin{align}
    \psi(t,k) &\,\simeq\, \chi_0(k) \,e^{-\left(\tfrac12-m\right)t} \hat \psi_+ +\mathcal O(k) \,e^{-\left(\tfrac12+m\right)t} \hat \psi_-,\\
    \bar\psi(t,k) &\,\simeq\, \bar\chi_0(k) \,e^{-\left(\tfrac12-m\right)t} \hat \psi_- +\mathcal{\bar{O}}(k) \,e^{-\left(\tfrac12+m\right)t} \hat \psi_+.
\end{align}
The spinors are defined by $\hat \psi_\pm = (1,\,\pm1)^t$ with the boundary field value of $\chi_0$. The boundary term in the action is evaluated and the remaining contribution of LO in the limit of $t\to\infty$ is
\begin{equation}
    \frac12 \int d\theta\,a(t) \bar\psi(t,\theta) \gamma^0 \psi(t,\theta) \bigg|_{t\to \infty} \,=\, i\sum_k \left[\bar\chi_0(-k) \mathcal{ O}(k) - \mathcal{\bar O}(-k) \chi_0(k) \right].
\end{equation}
In the wormhole geometry, the field has the boundary values of $\chi_{1,2}$ at each boundary. The quantities of $\mathcal O$ are written schematically by $\mathcal O_1 = a\chi_1 +b\chi_2,\,\mathcal O_2 = b\chi_1 - a\chi_2$. The boundary action would read
\begin{align}
    iS_b &\,=\, \frac i2 \left( \int d\theta\,a(t) \bar\psi_1(t,\theta) \gamma^0 \psi_1(t,\theta) -\int d\theta\,a(t) \bar\psi_2(t,\theta) \gamma^0 \psi_2(t,\theta)  \right) \\ \nonumber
    &\,=\, -2 \sum_{k\in \mathbb Z+\tfrac12} \left[a_k \bar\chi_2(-k)\chi_2(k)  - b_k \left(\bar\chi_2(-k)\chi_1(k)+\chi_2(k)\bar\chi_1(-k)\right)+a_k \bar\chi_1(-k) \chi_1(k) \right].
\end{align}
The $\chi_2$ are integrated out as they are in the unobservable boundary and it provides the modified action yielding the probability and the correlator,
\begin{equation}
    \rho[\chi_1,\chi_1] \,=\, \mathcal N \int D\bar\chi_2 D\chi_2\,\exp \left[i (S_b-S_b^*) \right] \,\sim\, \prod_{k\in \mathbb Z+\tfrac12}\exp \left[-4\left({\rm Re} a_k + \frac{({\rm Im}b_k)^2}{{\rm Re }a_k} \right) \bar\chi_1(-k) \chi_1(k) \right].
\end{equation}
The fermionic contribution to the partition function depends on the product of the quadratic coefficients, rather than on their inverse. As a consistency, the coefficient of $\bar\chi \chi$ has the correction sign in order that the partition function is positive and has the appropriate physical meaning.

\begin{figure}[t]
    \centering
    \includegraphics[width=0.5\linewidth]{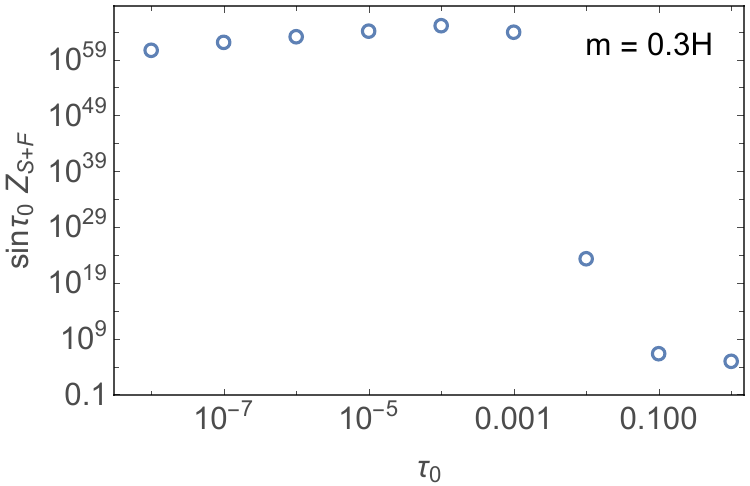}
    \caption{This figure shows the $Z_{rat}$ , including the two scalar fields and a spinor field with $k_{\rm max} =200$.  Fermions and scalars compete to yield nonzero finite saddle point of $\tau_0$, typically $\sim 10^{-3}-10^{-2}$ for the given parameters of the universe.}
    \label{Fig:ZratSF03}
\end{figure}

This coefficient vanishes linearly in $\tau_0 \to 0$, and therefore compensates the $1/\tau_0$-type divergence arising from the scalar sector. If the particle spectrum consists of two scalar fields and one fermion, the combined matter partition function can be written schematically as
\begin{equation}
    Z^{WH} \,=\, \int D\phi D\bar\chi D\chi \,\big|\Psi[\phi,\chi]\big|^2 \,=\, \prod_k \frac{1}{\,{\rm Im}A_k^{\rm scalar}}\,\cdot\left({\rm Re} a_k + \frac{({\rm Im}b_k)^2}{{\rm Re }a_k} \right),
\end{equation}
where $A_k^{\rm scalar}$ is the scalar kernel which vanishes linearly in $\tau_0$. One can check that the total integrand remains finite even in the limit of $\tau_0 \to 0$.

We can then compute the ratio $Z_{S+F}=Z^{WH}(\tau_0)/Z^{HH}$ for this combined matter spectrum $(S+F)$. As shown in \Fig{Fig:ZratSF03}, the resulting partition function is finite throughout the range of $\tau_0$, and the dominant contribution comes from a saddle at nonzero $\tau_0$.

\section{Suppression of contributions from multiple wormholes}\label{app:Magnet}
We argue that the tracing-out procedure allows multiboundary wormholes to contribute to two-boundary quantities in the path integral. In particular, one may consider the possibility that geometries with more than two boundaries can be traced out to yield an two-boundary quantity. In this appendix we analyze this possibility and conclude that their contributions are suppressed by the number of wormhole geometries. 

\begin{figure}[t]
    \centering
    \includegraphics[width=0.6\linewidth]{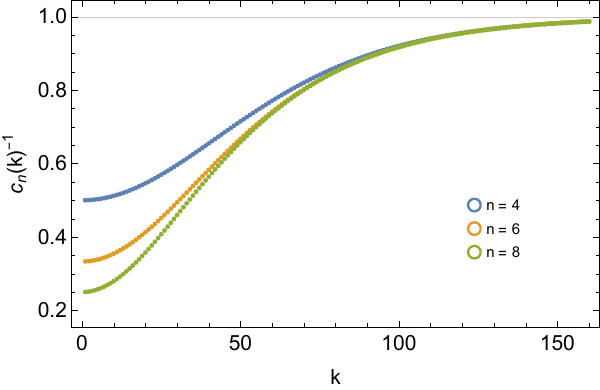}
    \caption{Inverse ratio factor $c_n(k)^{-1}$ to the leading coefficients for traced multi-wormholes. The three colored curves correspond to an increasing number of traced wormholes for $n=4,6,8$. For small $k$ the factor lies well below~$1$, indicating a strong suppression when all $c(k)^{-1}$ are multiplied.}
    \label{fig:MBKcnk}
\end{figure}

We can connect more than two wormholes and trace them out systematically to obtain the wave functional. The resulting wormhole geometry produces an amplitude of the very schematic form
\begin{equation}
  iS[\phi_1,\phi_2]\;=\;-a\,\phi_1^{2}\,+\,2b\,\phi_1\phi_2\,-\,a\,\phi_2^{2}.
\end{equation}
Here, $\phi_{1,2}$ are the boundary values of the field on the two boundaries the wormhole saddle connects. For the main results' case $(n=2)$, we have
\begin{equation}
  \mathcal N \int \mathcal D\phi_0\;
  \exp\!\bigl[i\,S[\phi_0,\phi]-i\,S^{*}[\phi_0,\phi]\bigr]
  \;\sim\;
  \exp\!\bigl(-\mathcal I \,\phi^{2}\bigr),
\end{equation}
where $\mathcal I\equiv (\alpha^2-\beta^2)/\alpha$. The notations are same with ones in the \ref{sec:trace_univ}.
The $n=4$ contribution is obtained by tracing out three boundary values $(\phi_{0,1,2})$ and is evaluated as
\begin{equation}
  \mathcal N \int \mathcal D\phi_0\,\mathcal D\phi_1\,\mathcal D\phi_2\;
  \exp\!\bigl[i\,S[\phi_0,\phi_1]-i\,S^{*}[\phi_0,\phi_2]
             +i\,S[\phi_2,\phi]-i\,S^{*}[\phi_1,\phi]\bigr]
  \;\sim\;
  \exp\!\bigl(-\mathcal I\,c_{n=4}\,\phi^{2}\bigr).
\end{equation}
Here $c_n$ denotes the ratio of the $n$-wormhole contribution to that of the leading bra--ket wormhole. Tracing out four wormholes therefore yields a probability distribution only for $\phi$. Comparing the various contributions, the partition function obtained after integrating over $\phi$ becomes
\begin{equation}
  \int \exp\!\left[-\sum_k \mathcal I(k)\,c_n(k)\,\phi_k^{2}\right]
  \;=\;
  \prod_k \sqrt{\frac{\pi}{c_n(k)\,\mathcal I(k)}}
  \;=\;
  \left(\prod_k \frac{1}{\sqrt{c_n(k)}}\right)\,Z_{n=2}.
\end{equation}
Hence the overall normalization from $n$ wormholes is suppressed by the product of factors $1/\sqrt{c_n(k)}$, as shown in \Fig{fig:MBKcnk}. Each factor satisfies $0<c_n(k)<1$ and decreases further as $n$ becomes large or $\tau_0$ becomes smaller. It can be probed that the $c_n(k=0)\propto n $ by the mathematical induction. Consequently, the more wormholes are traced out, the more strongly their contributions are suppressed in the path integral.

\bibliographystyle{JHEP}
\bibliography{ref.bib}
\end{document}